\documentclass{article}

\usepackage{arxiv}

\usepackage[utf8]{inputenc} % allow utf-8 input
\usepackage[T1]{fontenc}    % use 8-bit T1 fonts
\usepackage{hyperref}       % hyperlinks
\usepackage{url}            % simple URL typesetting
\usepackage{booktabs}       % professional-quality tables
\usepackage{amsfonts}       % blackboard math symbols
\usepackage{nicefrac}       % compact symbols for 1/2, etc.
\usepackage{microtype}      % microtypography
\usepackage{graphicx}
\usepackage{natbib}
\usepackage{doi}

\usepackage{xcolor}
\hypersetup{
  colorlinks=true,
  linkcolor=red,
  citecolor=orange,
  urlcolor=blue,
}

\graphicspath{{../figure}}

\newcommand\Rey{\mbox{\textit{Re}}}  % Reynolds number
\newcommand\Pran{\mbox{\textit{Pr}}} % Prandtl number, cf TeX's \Pr product

\title{Interference and heat transfer between hairpin vortices in wakes behind staggered hills
}

\author{\href{https://orcid.org/0000-0002-4875-8174}{\includegraphics[scale=0.06]{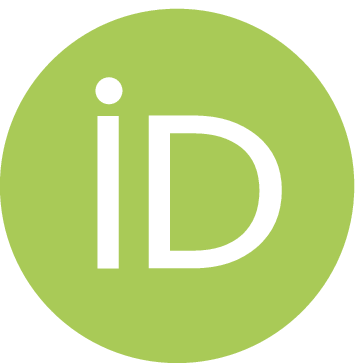}\hspace{1mm}Hideki Yanaoka}
\thanks{Email address for correspondence: yanaoka@iwate-u.ac.jp} \\
	Department of Systems Innovation Engineering, \\
    Faculty of Science and Engineering, Iwate University, \\
    4-3-5 Ueda, Morioka1, Iwate 202-8551, Japan \\
	\And
	Yoshiyuki Yomogida \\
	SUBARU CORPORATION, \\
	1-20-8 Ebisu, Shibuya-ku, Tokyo 150-8554, Japan
	\texttt{} \\
}

\renewcommand{\shorttitle}{Interference and heat transfer between hairpin vortices in wakes behind staggered hills}

\makeatletter
\hypersetup{
pdfusetitle,
bookmarksnumbered,
pdftitle={\@title},
pdfsubject={},
pdfauthor={Hideki Yanaoka, Yoshiyuki Yomogida},
pdfkeywords={Vortex dynamics, Hairpin vortex, Heat transfer, Unsteady flow, Boundary layer separation, Numerical simulation},
}
\makeatother

\begin{document}

\maketitle

% Abstract
\begin{abstract}
The present study performs a numerical simulation of the interference 
and heat transfer between hairpin vortices formed in wakes 
behind staggered hills in a laminar boundary layer. 
Hairpin vortices are periodically shed in the wake of a row of hills, 
causing interference between the hairpin vortices. 
As the spanwise distance between the hills decreases, 
interference increases and the hairpin vortices become strong. 
At that time, because the interference between the legs of the hairpin vortex 
and the Q2 ejection becomes strong, 
the head of each hairpin vortex rises sharply. 
When the hill spacing decreases, 
the turbulence caused by the head and both legs of the hairpin vortex 
generated from a hill in the second row increases remarkably. 
In addition, the secondary vortex also generates turbulence. 
The hairpin vortex and the secondary vortex are attracted 
to adjacent hairpin vortices, 
causing widespread high turbulence in the spanwise direction 
near the wall surface. 
Regardless of the hill spacing, 
Q2 ejection and Q4 sweep due to the hairpin vortex occur, 
and the secondary vortex forms around the hairpin vortex, 
activating heat transport and increasing the heat transfer coefficient 
in the wake. 
When the hill spacing becomes narrower, 
the interference between the hairpin vortices strengthens 
the legs of each hairpin vortex and secondary vortex, 
and heat transport near the wall surface becomes very active. 
The heat transfer increases over a wide range of the wake 
because the legs of hairpin vortices flowing downstream are spread 
in the spanwise direction.
\end{abstract}

% Keywords

%##############################################################################
\section{Introduction}
%##############################################################################

Hairpin vortices are a typical coherent vortex structure present 
in the boundary layer \citep{Acarlar&Smith_1987a}. 
The presence of multiple hairpin vortices has been observed 
in the boundary layer transition process and in a turbulent boundary layer 
\citep{Christensen&Adrian_2001, Bake_et_al_2002, Green_et_al_2007, 
Bernard_2011, Lu_et_al_2013, Schlatter_et_al_2014}. 
Detailed research on the characteristics of hairpin vortices 
has been conducted to understand the complicated flow field 
\citep{Acarlar&Smith_1987a, Gretta&Smith_1993, Zhou_et_al_1999, 
Yang_et_al_2001, Dong&Meng_2004, Adrian_2007, Yanaoka_et_al_2007b, Yanaoka_et_al_2008b, Li_et_al_2019, Lloyd_et_al_2022}.

Hairpin vortices, which constitute a vortex structure with longitudinal vortices called legs, 
develop along a wall surface. 
Previous studies \citep{Torii_et_al_1994, Senaha_et_al_2001} 
have shown that such longitudinal vortices are less likely to decay, 
thus facilitating fluid mixing and heat transfer in the wake. 
\citet{Acarlar&Smith_1987a} investigated a hairpin vortex 
in the wake of a hemispherical protuberance in a visualization experiment 
and found that both legs of the hairpin vortex extend 
in the downstream direction. 
The present authors \citep{Yanaoka_et_al_2007b,Yanaoka_et_al_2008b} previously reported 
that hairpin vortices cause high turbulence in the wake 
and increase heat transfer, as determined from a three-dimensional numerical analysis 
of a hairpin vortex generated behind a hill. 
\citet{Li_et_al_2019} investigated the coherent structure and heat transfer 
in the turbulent boundary layer along the wall of a channel with a rib tabulator 
and found that multiple hairpin vortices occur, 
increasing the heat transfer coefficient and turbulence. 
Therefore, the formation of hairpin vortices is expected 
to be an effective method for mixing substances 
and promoting heat transfer. 

When multiple ribs are installed on a wall, 
the wakes behind the ribs interfere with each other. 
Therefore, the vortex characteristics in a wake 
change depending on the arrangement of the ribs. 
\citet{Kurita&Yahagi_2008} investigated the wake of heated cylinders 
arranged perpendicular to the flow direction. 
They found that the wake becomes symmetric (asymmetric) when the cylinder spacing 
is wide (narrow). 
\citet{Hanson_et_al_2009} showed 
that when two cylinders are placed perpendicular to the flow direction, 
there are two vortex shedding frequencies 
when the distance between the cylinders is narrow. 
Studies on such wake interference 
have not targeted flow fields with multiple hairpin vortices 
\citep{Moriya&Sakamoto_1985, Sayers&Saban_1994, Zhou_et_al_2002, 
Wang&Zhou_2005, Li&Sumner_2009}. 
In a complex flow field with multiple hairpin vortices 
that generate high turbulence, it is considered that the turbulence 
and heat transfer change significantly due to the interference 
between the hairpin vortices.

In this study, we perform a numerical analysis of a flow field 
where multiple hairpin vortices exist in the wake 
behind staggered hills in a laminar boundary layer 
and clarify the effects of the interference between hairpin vortices 
on turbulence and heat transfer.

%##############################################################################
\section{Fundamental equation and numerical procedures}
%##############################################################################

\begin{figure}[t!]
\begin{center}
\includegraphics[trim=0mm 0mm 0mm 0mm, clip, height=25mm]{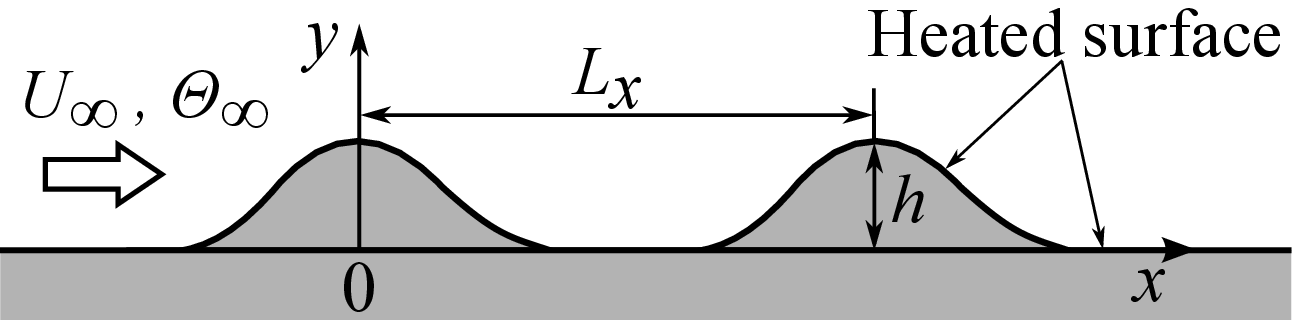} \\
(a) $x$-$y$ plane \\
\vspace*{1.0\baselineskip}
\includegraphics[trim=0mm 0mm 0mm 0mm, clip, height=25mm]{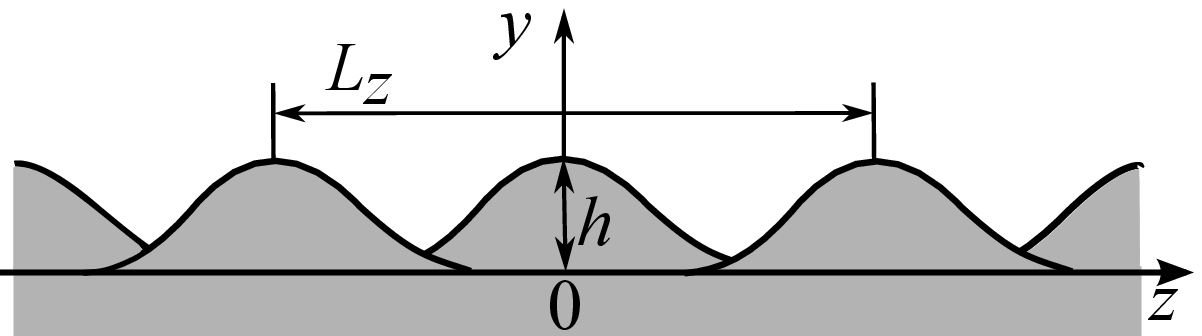} \\
(b) $y$-$z$ plane   
\end{center}
\caption{Flow configuration and coordinate system.}
\label{flow_model1}
\end{figure}

Figure \ref{flow_model1} shows the flow configuration and coordinate system. 
The origin is on the wall surface and the $x$-, $y$-, and $z$-axes are 
the streamwise, cross-streamwise, and spanwise directions, respectively. 
The velocities in these directions are denoted as $u$, $v$, and $w$, respectively, 
and the temperature is denoted as $\theta$. 
A uniform flow originates upstream and a laminar boundary layer develops 
downstream. 
The free stream velocity and temperature are denoted as
$U_{\infty}$ and $\Theta_{\infty}$, respectively. 
In this study, similar to \citet{Acarlar&Smith_1987a}, 
a rib is placed in the laminar boundary layer 
and a hairpin vortex is generated using flow separation. 
The rib has the same shape as that used in the experiment of 
\citet{Simpson_et_al_2002}; it is a hill with a height of $h$ 
and a radius of $2h$. 
The height of the hill is given as follows:
%------------------------------------------------------------------------------
\begin{equation}
  \frac{y(r)}{h} = - \frac{1}{6.04844}\biggl[J_0(\Lambda)I_0
  \biggl(\Lambda\frac{r}{a} \biggl) - I_0(\Lambda)J_0
  \biggl(\Lambda\frac{r}{a} \biggl) \biggl], 
  \label{hill_defined}
\end{equation}
%------------------------------------------------------------------------------
where $\Lambda=3.1926$, $r$ is the radius, $h$ is the hill height, 
$a=2h$ is the hill radius, $J_0$ is the Bessel function of the first kind, 
and $I_0$ is the modified Bessel function of the first kind.

%------------------------------------------------------------------------------
% Figure 1
%------------------------------------------------------------------------------
%------------------------------------------------------------------------------

In this study, to create a flow field with multiple hairpin vortices, 
hill-shaped ribs are arranged parallel to the spanwise direction 
on the wall surface, as shown in figure \ref{flow_model1}, 
and installed in a staggered pattern. 
The hill spacing in the streamwise direction is $L_x$ 
and that in the spanwise direction is $L_z$. 
The spanwise distance between the vertices of the hills in the first 
and second rows is $L_z/2$.

This study deals with the three-dimensional flow 
of an incompressible viscous fluid with constant physical properties. 
The continuity, Navier-Stokes, and energy equations 
in Cartesian coordinates are transformed into those in a general coordinate system. 
The governing equations are solved using the simplified marker and cell method 
\citep{Amsden&Harlow_1970} extended to the collocated grid. 
The pressure interpolation proposed by \citet{Rhie&Chow_1983} 
is used to remove spurious errors. 
The Crank-Nicolson method is applied to discretize the time derivative 
and then time marching is performed. 
The second-order central difference scheme is used to discretize 
the spatial derivative.

%##############################################################################
\section{Details of calculation}
%##############################################################################

In this study, the hill spacing in the streamwise direction is fixed 
at $L_x=6h$ 
and the spanwise hill spacing is $L_z=4h$ or $7h$. 
We vary $L_z$ to investigate the effect of interference 
between hairpin vortices on the flow field and heat transfer. 
The computational region is $-10h$ to $40h$ in the $x$-direction, 
0 to $8h$ in the $y$-direction, and $-3.5h$ to $3.5h$ ($-2h$ to $2h$) 
for $L_z=7h$ ($4h$) in the $z$-direction.

For the boundary conditions of the velocity field, 
the Blasius velocity profile is given at the inlet 
and convective boundary conditions are used at the outlet. 
No-slip boundary conditions are applied on the walls. 
Slip boundary conditions are assumed at the upper boundary 
away from the bottom surface. 
For the boundary condition of the temperature field, 
a uniform temperature is given at the inlet 
and the convective boundary condition is used at the outlet. 
The hill surface and bottom wall are heated with a constant heat flux. 
A zero gradient of the temperature is assumed at the upper boundary. 
Periodic boundary conditions are applied in the spanwise direction 
for the velocity and temperature.

The numerical calculations are performed under the Reynolds number $\Rey=500$, 
which is defined by $h$ and $U_{\infty}$, and the Prandtl number $\Pran=0.7$. 
The velocity boundary layer thickness $\delta$ at the inlet 
is set so that the thickness becomes $\delta_0/h=1$ at $x/h=0$ 
without the hill. 
The boundary layer thickness at the entrance is $\delta/h=0.707$, 
the displacement thickness is $\delta_d/h=0.243$, 
and the momentum thickness is $\delta_m/h=0.094$.
The Reynolds numbers defined by these thicknesses and $U_{\infty}$ are 
$\Rey_{\delta}=354$, $\Rey_{\delta_d}=122$, and $\Rey_{\delta_m}=47$, respectively. 
These conditions are the same as those in previously reported calculations 
\citep{Yanaoka_et_al_2007b, Yanaoka_et_al_2008b}.

Grids with dimensions of $426 \times 81 \times 83$ (grid7-1), 
$463 \times 91 \times 95$ (grid7-2), and $478 \times 101 \times 105$ (grid7-3) 
are used to confirm the grid dependency on the calculation result for $L_z=7h$. 
Grids with dimensions of $381 \times 81 \times 63$ (grid4-1), 
$426 \times 94 \times 75$ (grid4-2), $478 \times 101 \times 83$ (grid4-3), 
and $531 \times 106 \times 89$ (grid4-4) are used for $L_z=4h$. 
The grids are dense near the lower wall surface and hills. 
The minimum grid widths for $L_z=7h$ are $0.01h$ (grid7-1), $0.005h$ (grid7-2), 
and $0.0025h$ (grid7-3).
The minimum grid widths for $L_z=4h$ are $0.01h$ (grid4-1), $0.0075h$ (grid4-2), 
$0.005h$ (grid4-3), and $0.0025h$ (grid4-4). 
For each $L_z$ value, we investigated the effect of the number of grid points 
on the calculation results. 
We confirmed that the differences between the results obtained with grid7-2 and grid7-3 
for $L_z=7h$ and between those obtained with grid4-3 and grid4-4 for $L_z=4h$ were small. 
The results obtained using the fine grids, namely grid3 for $L_z=7h$ 
and grid4 for $L_z=4h$, are shown to clarify 
the turbulence due to the vortex structure in detail.

To clearly show the interference between hairpin vortices in the wake of a hill, 
we expanded the visualization region in the spanwise direction 
to visualize the areas from $-7h$ to $7h$ for $L_z=7h$ and from $-4h$ to $4h$ for $L_z=4h$.

%------------------------------------------------------------------------------
% Figure 2
%------------------------------------------------------------------------------
\begin{figure}
\begin{center}
\includegraphics[trim=3mm 0mm 0mm 0mm, clip, width=100mm]{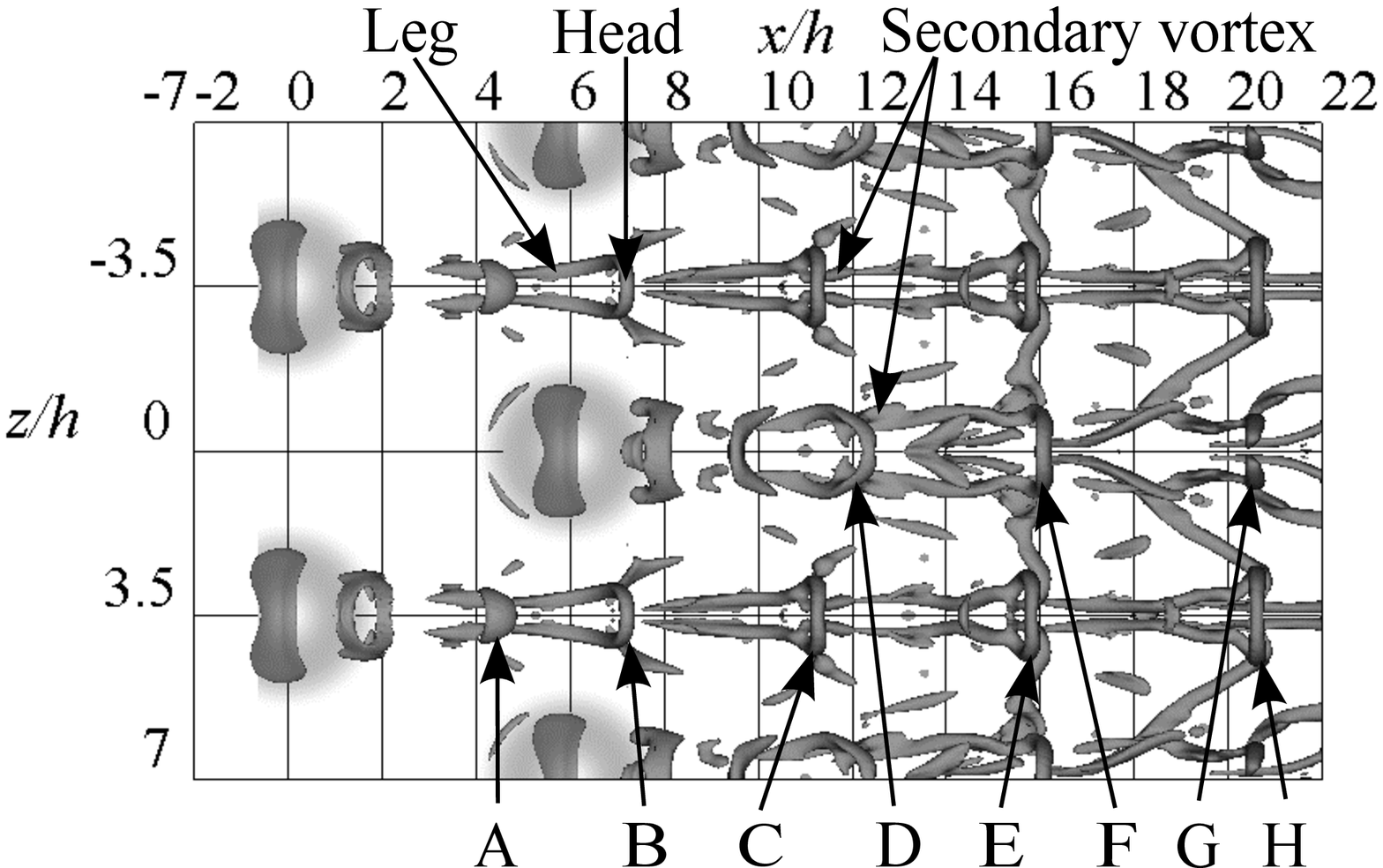} \\
(a) Top view \\
\vspace*{0.5\baselineskip}
\includegraphics[trim=0mm 0mm 1mm 4mm, clip, width=90mm]{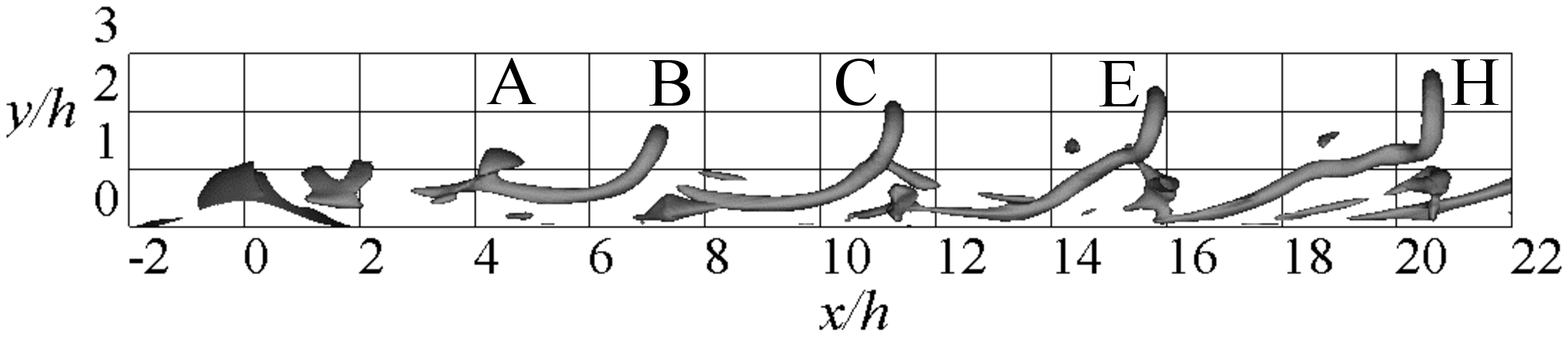} \\
(b) Side view downstream of hill in first row \\
\vspace*{0.5\baselineskip}
\includegraphics[trim=0mm 0mm 1mm 4mm, clip, width=90mm]{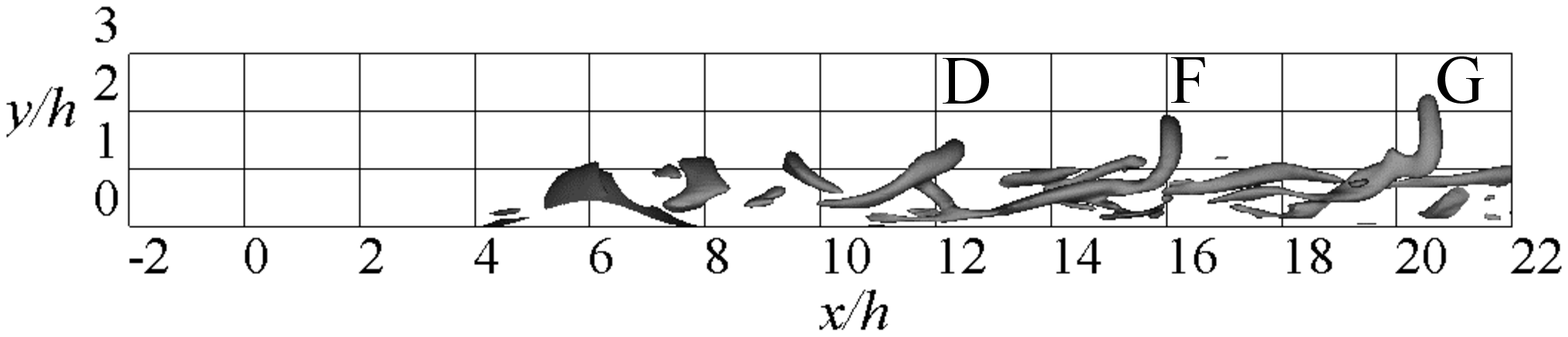} \\
(c) Side view downstream of hill in second row \\
\caption{Isosurface of the curvature of equipressure surface for $L_z=7h$ 
(isosurface value is $-7/h$).}
\label{curv_lz7}
\end{center}
\end{figure}
%------------------------------------------------------------------------------

%##############################################################################
\section{Results and discussion}
%##############################################################################

%++++++++++++++++++++++++++++++++++++++++++++++++++++++++++++++++++++++++++++++
\subsection{Vortex structure in wake}
%++++++++++++++++++++++++++++++++++++++++++++++++++++++++++++++++++++++++++++++

\begin{figure}
\begin{center}
\includegraphics[trim=3mm 0mm 0mm 0mm, clip, width=100mm]{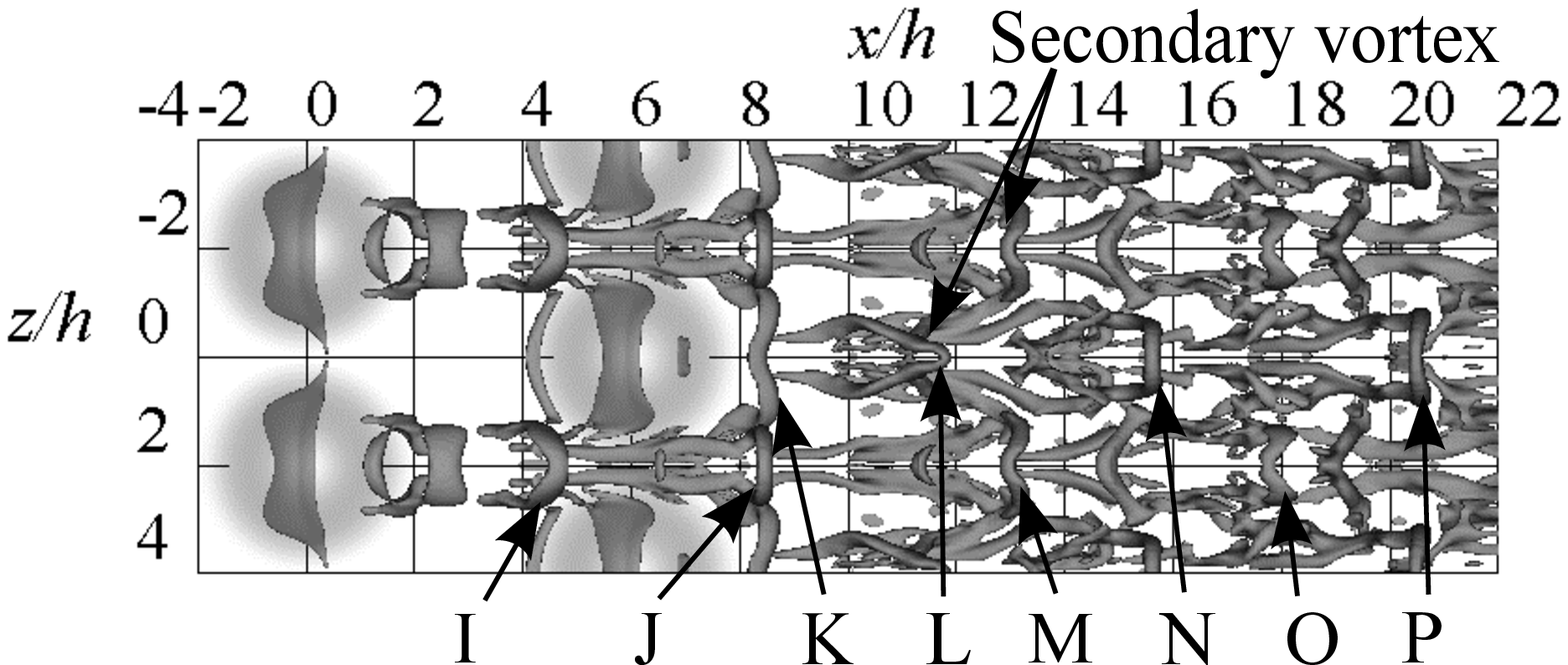} \\
(a) Top view \\
\vspace*{0.5\baselineskip}
\includegraphics[trim=0mm 0mm 1mm 4mm, clip, width=90mm]{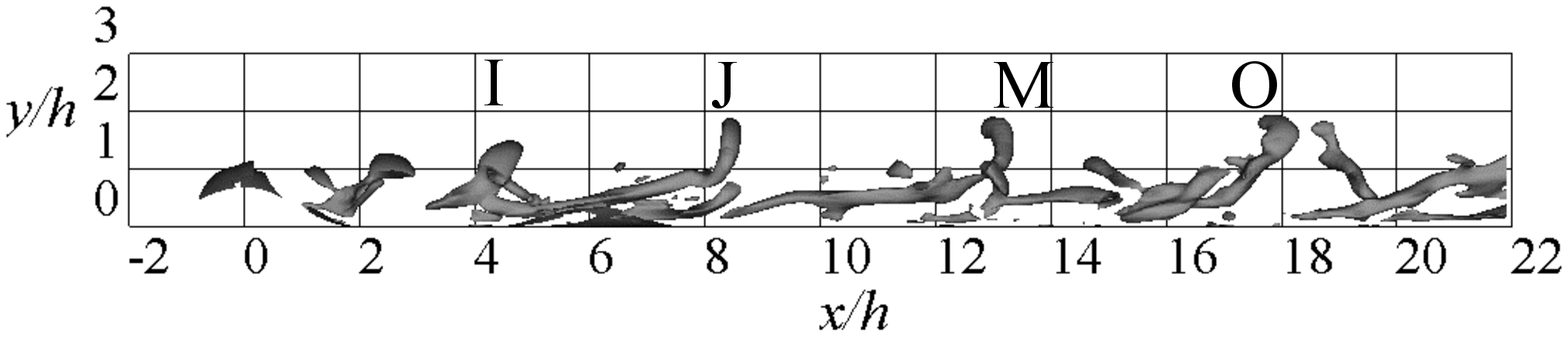} \\
(b) Side view downstream of hill in first row \\
\vspace*{0.50\baselineskip}
\includegraphics[trim=0mm 0mm 1mm 4mm, clip, width=90mm]{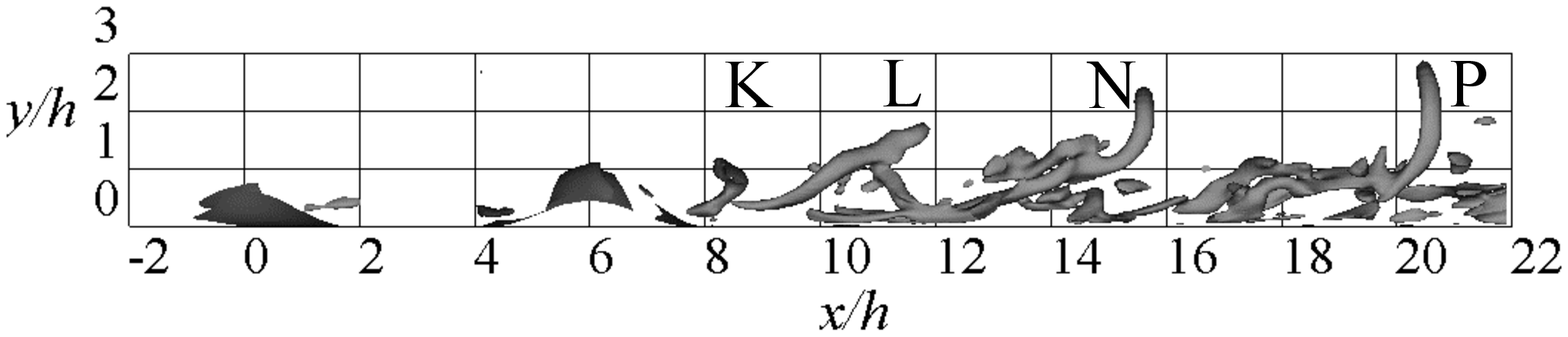} \\
(c) Side view downstream of hill in second row \\
\end{center}
\caption{Isosurface of curvature of equipressure surface for $L_z=4h$ 
(isosurface value is $-7/h$).}
\label{curv_lz4} 
\end{figure}

To clarify the vortex structure in the flow field, 
figures \ref{curv_lz7} and \ref{curv_lz4} show the top and side views of the isosurface 
of curvature calculated from the equipressure surface for $L_z=7h$ and $4h$, respectively. 
This method of visualizing the vortex structure was used in our previous research 
\citep{Yanaoka_et_al_2007b,Yanaoka_et_al_2007c,Yanaoka_et_al_2008b}.
In the figures, the hairpin vortices are labelled A to P. 
For $L_z=7h$ and $4h$, the separated shear layer at the top of each hill 
becomes unstable downstream and rolls up into a vortex. 
Such flow is similar to that for a single hill 
\citep{Yanaoka_et_al_2007b,Yanaoka_et_al_2008b}.

%------------------------------------------------------------------------------
% Figure 3
%------------------------------------------------------------------------------
%------------------------------------------------------------------------------

%------------------------------------------------------------------------------
% Figure 4
%------------------------------------------------------------------------------
\begin{figure}
\begin{minipage}{0.48\linewidth}
\begin{center}
\includegraphics[trim=0mm 0mm 0mm 0mm, clip, width=65mm]{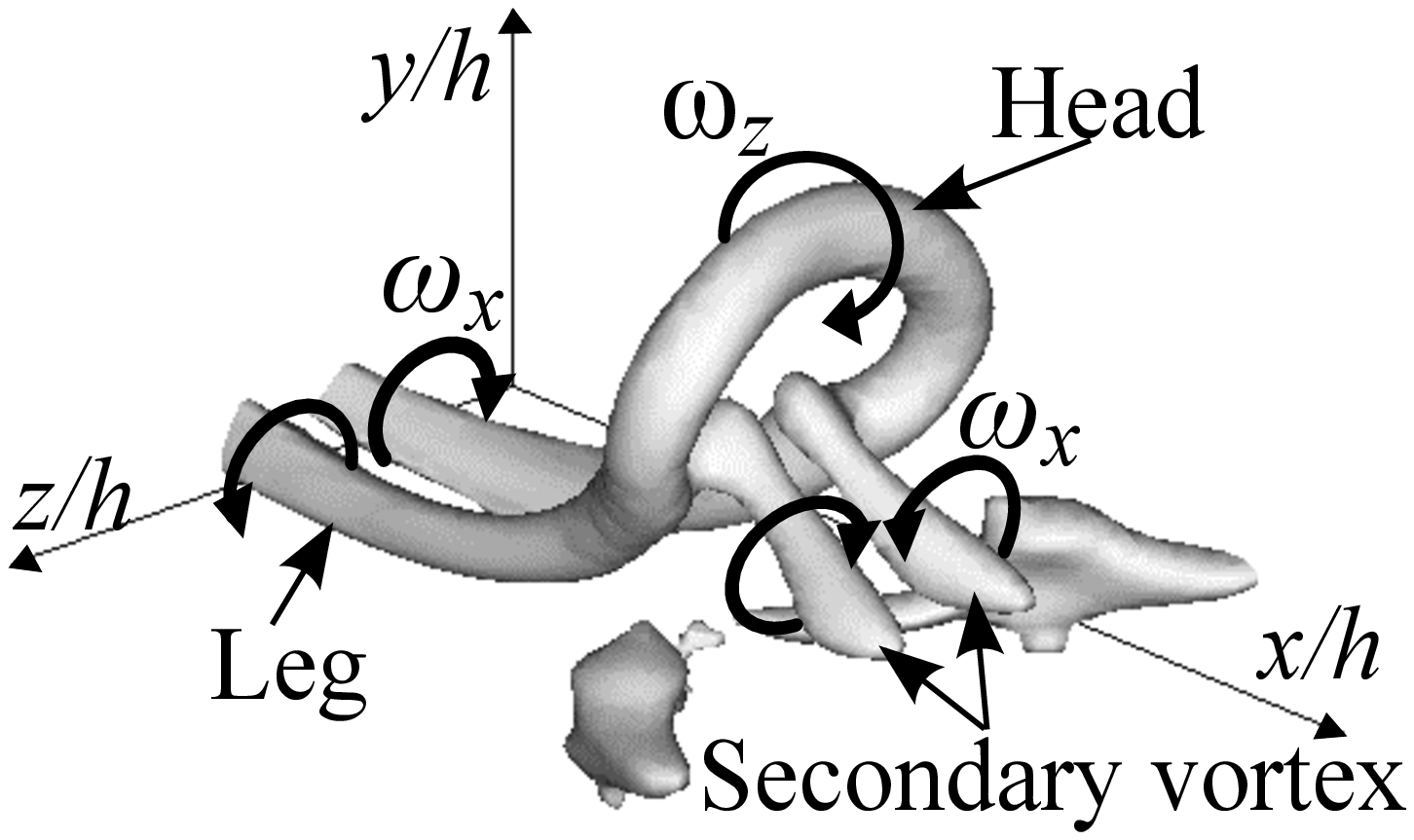} \\
(a) Perspective view
\end{center}
\end{minipage}
\hspace{0.02\linewidth}
\begin{minipage}{0.48\linewidth}
\begin{center}
\includegraphics[trim=0mm 0mm 0mm 0mm, clip, width=65mm]{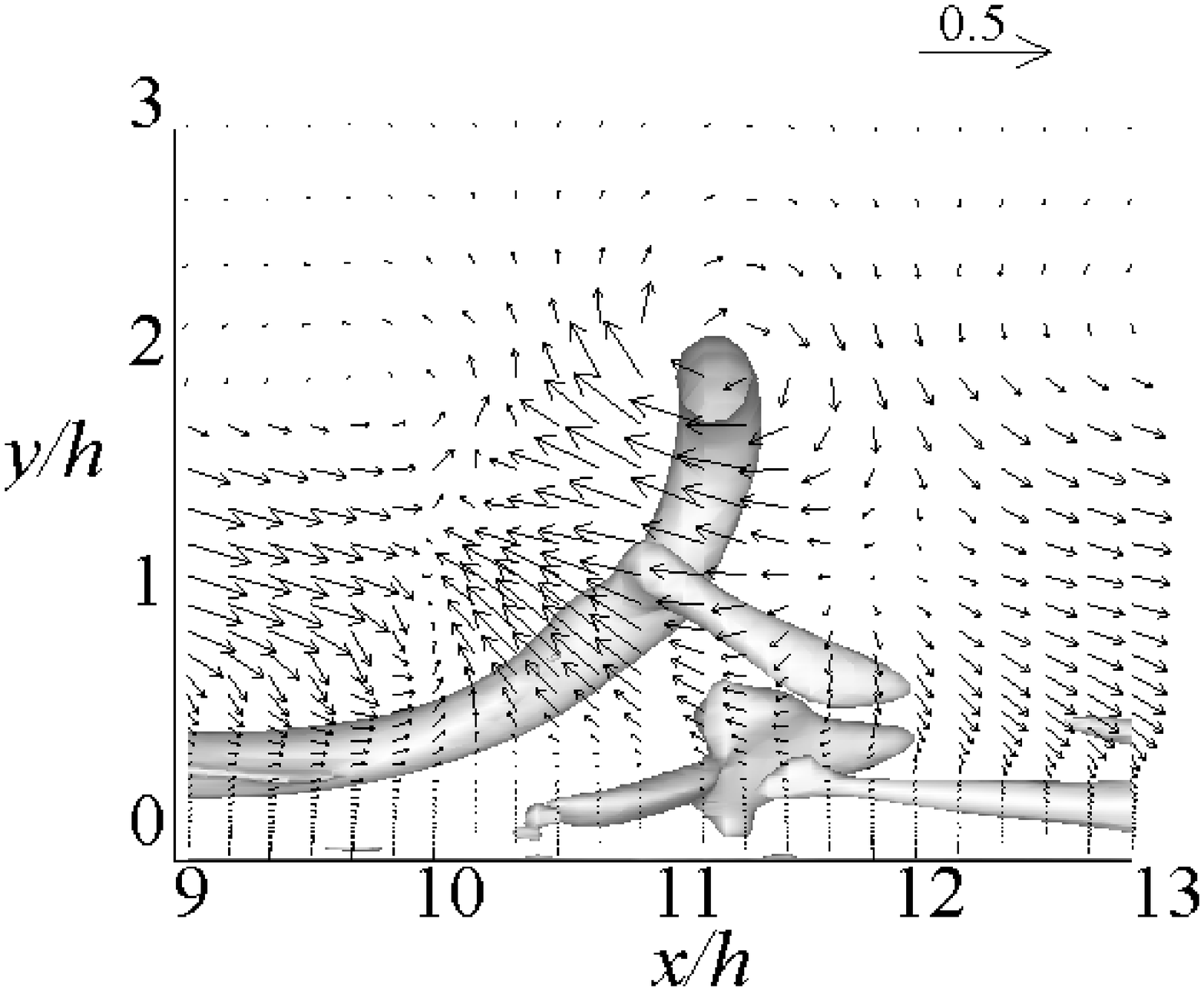} \\
(b) Velocity fluctuation vectors \\
\end{center}
\end{minipage}
\vspace*{1.0\baselineskip}
\begin{center}
\includegraphics[trim=0mm 0mm 0mm 0mm, clip, width=75mm]{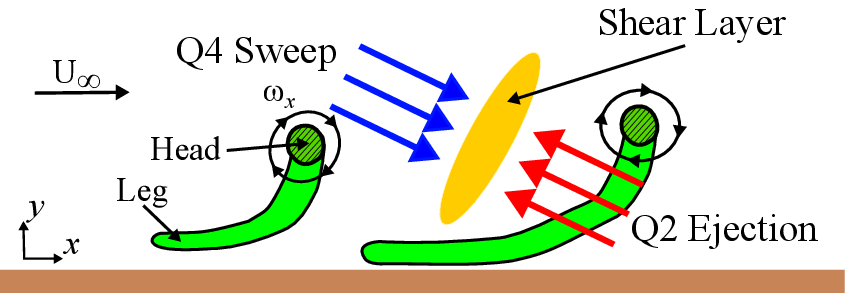} \\
(c) Schematic diagram \\
\end{center}
\caption{Dynamics of hairpin vortex C.} 
\label{induced}
\end{figure}
%------------------------------------------------------------------------------

For $L_z=7h$, shown in figure \ref{curv_lz7}(a), the vortices rolled up 
at $z/h=\pm 3.5$ and $z/h=0$ behind the hill rows grow into 
hairpin vortices A and D around $x/h=4$ and $x/h=2$ downstream, respectively. 
Under the heads of hairpin vortices C and D, a secondary vortex forms. 
This secondary vortex consists of two streamwise vortices toward the wall surface. 
As can be seen from figures \ref{curv_lz7}(b) and (c), as the hairpin vortex 
generated from each hill develops, the head rises 
and the angle between the legs and the wall increases. 
Because the shape and behaviour of such hairpin vortices are similar to 
those of a hairpin vortex generated from a single hill 
\citep{Yanaoka_et_al_2007b,Yanaoka_et_al_2008b}, 
it is considered that the interference between hairpin vortices is weak 
in this flow field.

For $L_z=4h$, shown in figure \ref{curv_lz4}(a), the vortex rolled up at $z/h=\pm 2$ 
behind the hills in the first row grows into hairpin vortex I 
around $x/h=4$. 
When the hairpin vortex moves downstream, 
its head shape changes in the same way as hairpin vortices J, M, and O. 
In addition, the vortex rolled up at $z/h=0$ behind the hill 
in the second row grows into hairpin vortex K around $x/h=8$. 
Downstream of hairpin vortex K, hairpin vortex L near $x/h=12$ 
deforms in the same way as hairpin vortices N and P. 
Below the heads of hairpin vortices L and M, 
secondary vortices are observed, as in the case of $L_z=7h$.
From figure \ref{curv_lz4}(b), it can be seen that the heads of 
hairpin vortices J, M, and O generated from the hills 
in the first row are at the same height. 
In contrast, as shown in figure \ref{curv_lz4}(c), 
the head of the hairpin vortex generated from the hill 
in the second row rises sharply downstream. 
The position of the head of hairpin vortex P reaches 
around $y/h=2.7$. 
The shape and behaviour of such hairpin vortices are very different 
from those of hairpin vortices generated from a single hill 
\citep{Yanaoka_et_al_2007b,Yanaoka_et_al_2008b}. 
Therefore, in the flow field for $L_z=4h$, 
it is considered that the hairpin vortices strongly interfere 
with each other.

Hairpin vortices generated for $L_z=7h$ and 4h are periodically shed 
in the wake behind each hill. 
The vortex shedding frequency is $f=0.18U_{\infty}/h$ for $L_z=7h$ 
and $f=0.17U_{\infty}/h$ for $L_z=4h$, 
indicating that it is mostly independent of hill spacing. 
These values are very close to the calculation result 
($f=0.16U_{\infty}/h$) for a single hill 
\citep{Yanaoka_et_al_2007b,Yanaoka_et_al_2008b}. 
Moreover, the calculation result is within the range of experimental results 
($f=0.13 U_{\infty}/h$ to $0.25U_{\infty}/h$) at $\Rey=500$ \citep{Acarlar&Smith_1987a}.

Figures \ref{induced}(a) and (b) show an enlarged view of hairpin vortex C 
in figure \ref{curv_lz7} and the velocity fluctuation vector in the $x$-$y$ plane, respectively, 
and figure \ref{induced}(c) shows the flow 
around the hairpin vortex. 
As shown in figure \ref{induced}(a), two streamwise vortices toward the wall surface 
appear below the head of hairpin vortex C. 
Such secondary vortices, which also exist 
under the heads of hairpin vortices D, L, and M, 
correspond to the horn-shaped secondary vortex reported 
in previous studies \citep{Yanaoka_et_al_2007b,Yanaoka_et_al_2008b}. 
In addition, it is found from figure \ref{induced}(b) that a flow ($\omega_z$) 
rotating clockwise is induced around the head of hairpin vortex C. 
The flow from the lower side of the hairpin vortex head 
to the upstream mainstream side is called Q2 ejection. 
The flow from the upstream of the hairpin vortex head 
to the downstream wall surface is called Q4 sweep (figure \ref{induced}(c)).
The classification of such flow phenomena was reported by 
\citet{Yang_et_al_2001}. 
Around $x/h=10$ in figure \ref{induced}(b), the flows due to Q2 ejection and Q4 sweep 
collide, generating a shear layer. 
The formation of such a shear layer has been previously observed 
\citep{Yang_et_al_2001, Yanaoka_et_al_2007b,Yanaoka_et_al_2008b}.

%------------------------------------------------------------------------------
% Figure 5
%------------------------------------------------------------------------------
\begin{figure}
\begin{center}
\includegraphics[trim=3mm 15mm 0mm 6mm, clip, width=90mm]{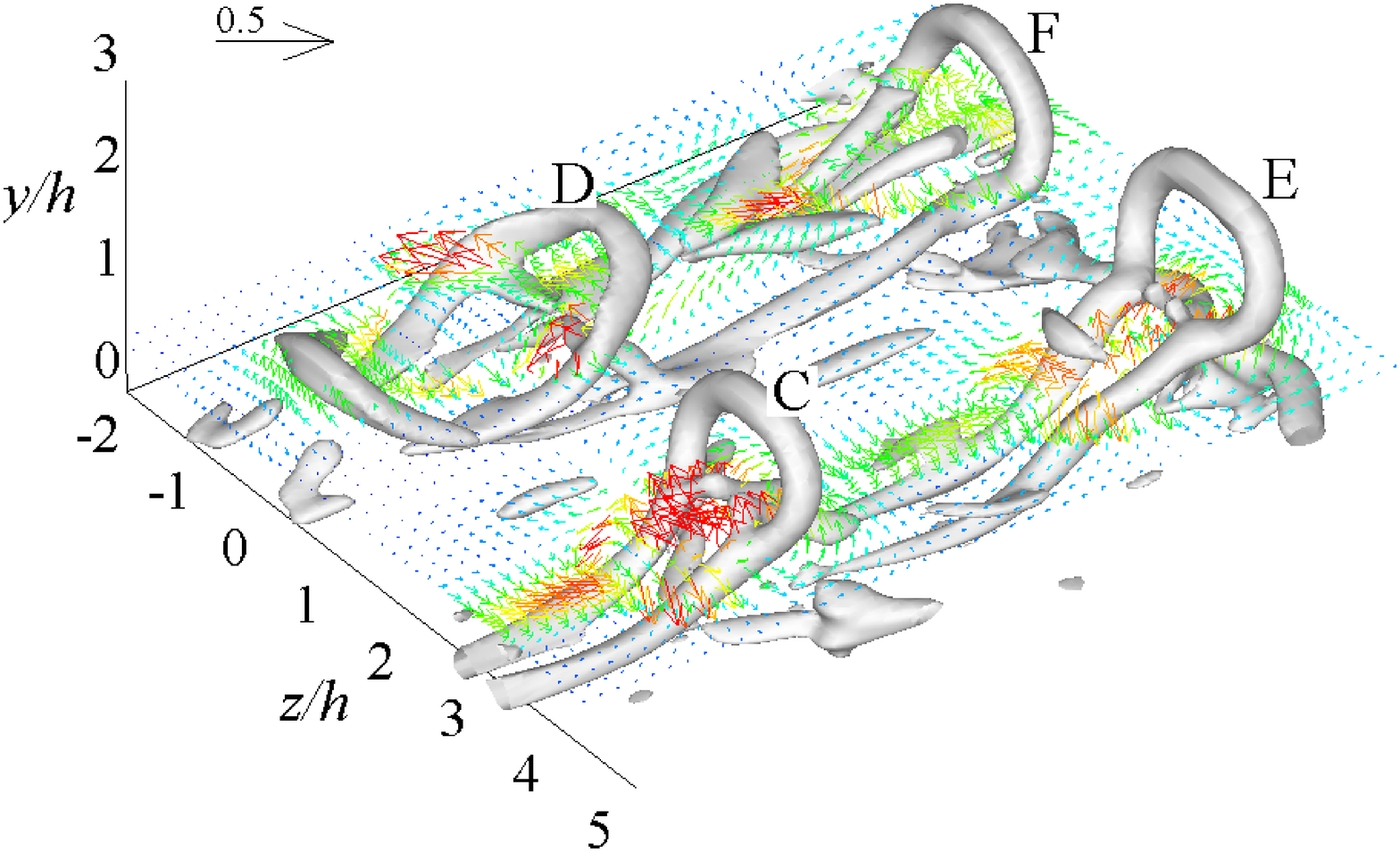} \\
(a) $L_z=7h$ \\
\vspace*{1.0\baselineskip}
\includegraphics[trim=3mm 3mm 0mm 2mm, clip, width=90mm]{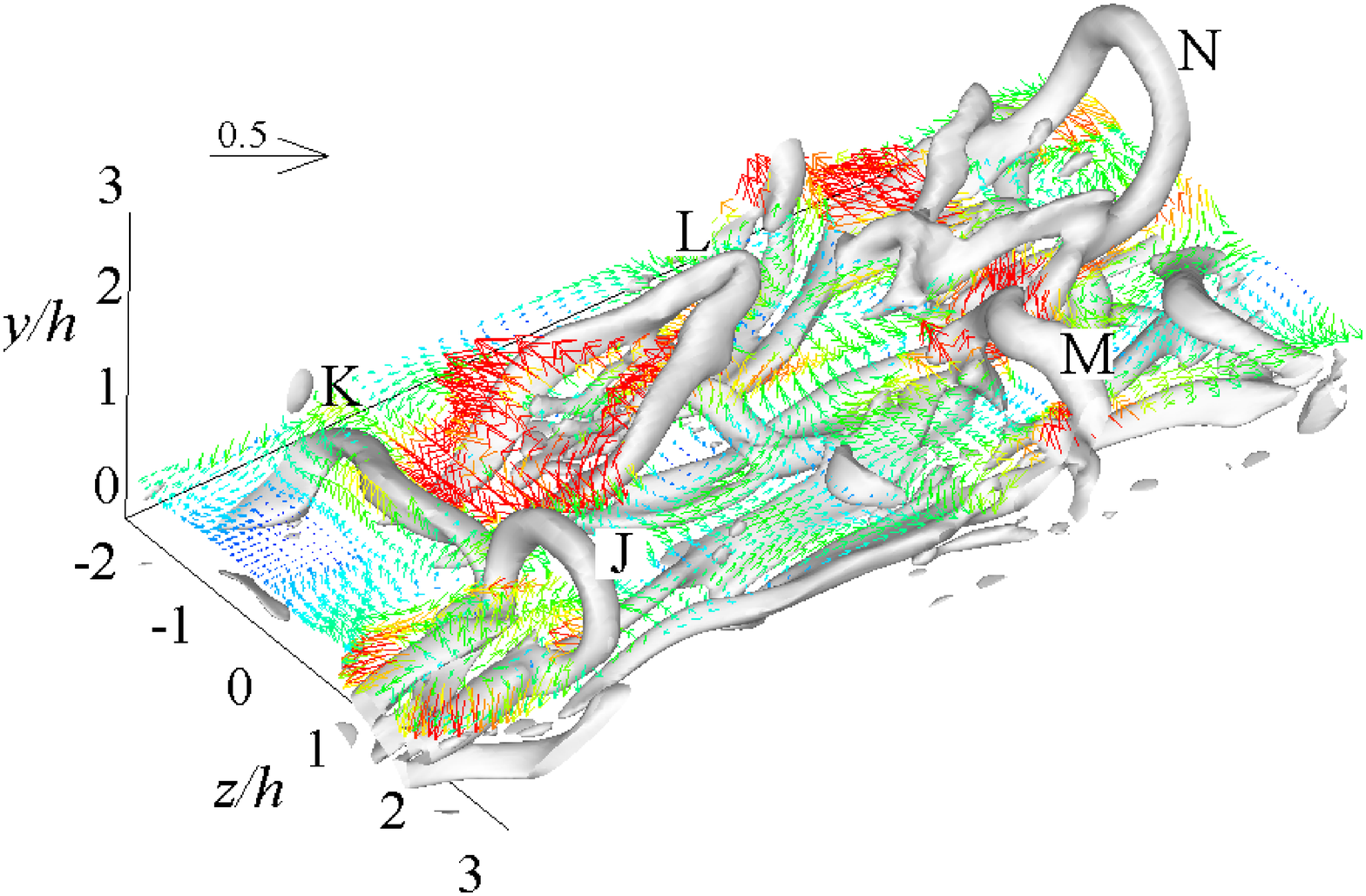} \\
(b) $L_z=4h$ \\
\end{center}
\caption{Velocity fluctuation vectors at $y/h=1$ around hairpin vortices (isosurface value is $-7/h$).}
\label{vector}
\end{figure}
%------------------------------------------------------------------------------

Next, to clarify the three-dimensional flow 
due to the hairpin vortices for $L_z=7h$ and $4h$, 
figure \ref{vector} shows the velocity fluctuation vector at $y/h=1$ 
and the isosurface of the curvature calculated 
from the equipressure surface. 
The position of $y/h=1$ is below the head of each hairpin vortex. 
As in figures \ref{curv_lz7} and \ref{curv_lz4}, the hairpin vortices 
in figure \ref{vector} are labelled C to F and J to M. 
The Q2 ejection observed in hairpin vortex C in figure \ref{induced}(b) 
and the ascending and descending flows due to the legs occur 
around all the hairpin vortices. 
For $L_z=4h$, the interference between hairpin vortices is stronger than that at
$L_z=7h$, and thus strong velocity fluctuations occur in the entire flow field 
and Q2 ejection due to the head of hairpin vortex L strengthens.

%------------------------------------------------------------------------------
% Figure 6
%------------------------------------------------------------------------------
\begin{figure}
\begin{minipage}{0.48\linewidth}
\begin{center}
\includegraphics[trim=1mm 2mm 2mm 2mm, clip, width=65mm]{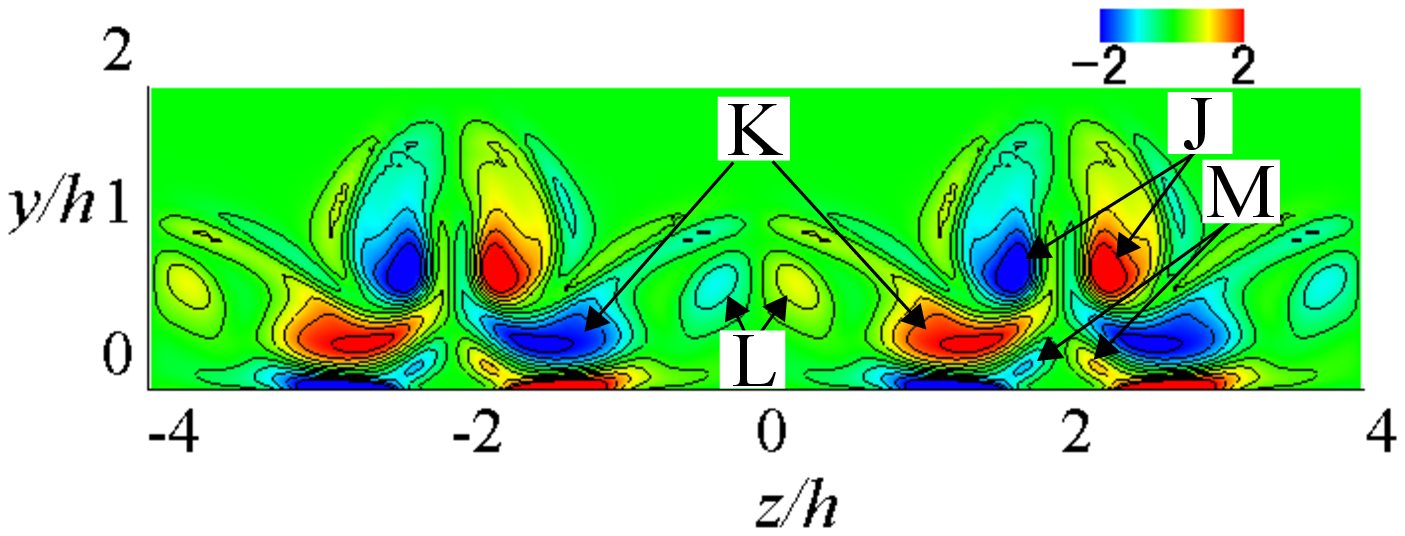} \\
(a) $x/h=8$
\end{center}
\end{minipage}
\hspace{0.02\linewidth}
\begin{minipage}{0.48\linewidth}
\begin{center}
\includegraphics[trim=1mm 2mm 2mm 2mm, clip, width=65mm]{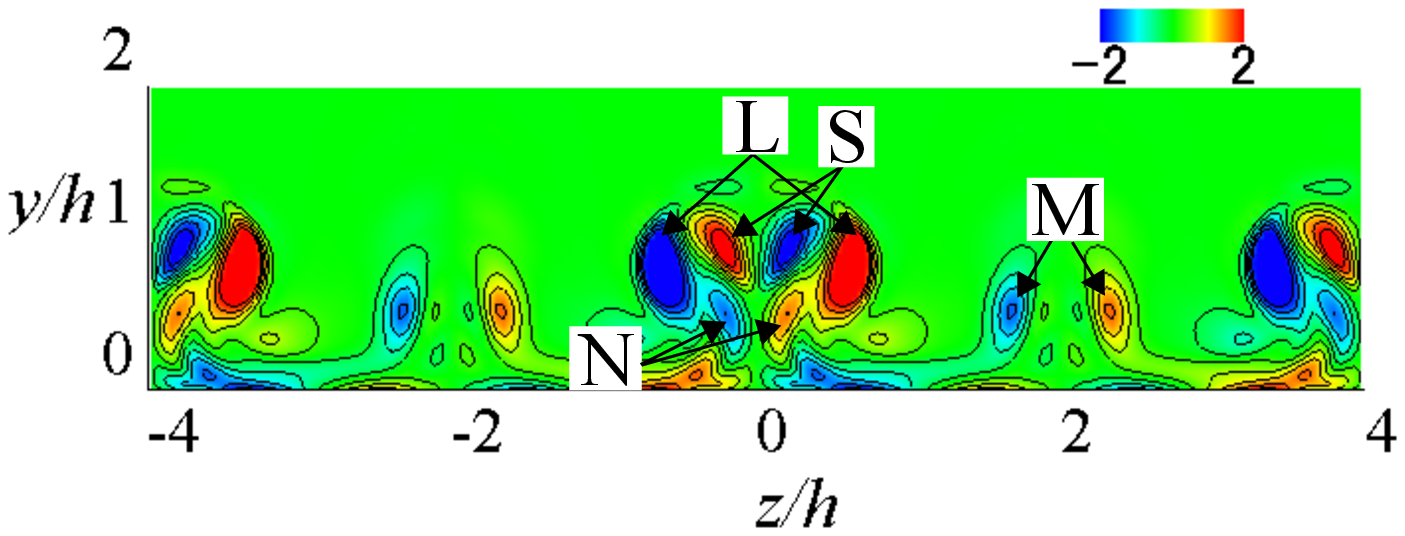} \\
(b) $x/h=10$ \\
\end{center}
\end{minipage}
\caption{Streamwise vorticity contours in $y$-$z$ plane for $L_z=4h$ 
(contour interval is 0.4 from $-2$ to 2).}
\label{vort_lz4}
\end{figure}
%------------------------------------------------------------------------------

%------------------------------------------------------------------------------
% Figure 7
%------------------------------------------------------------------------------
\begin{figure}
\begin{minipage}{0.48\linewidth}
\begin{center}
\includegraphics[trim=1mm 1mm 22mm 4mm, clip, width=65mm]{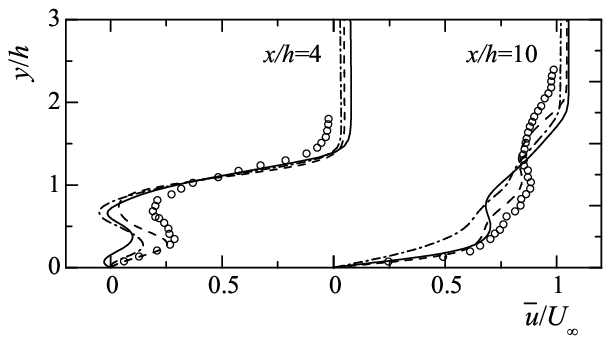} \\
(a) First hill
\end{center}
\end{minipage}
\hspace{0.02\linewidth}
\begin{minipage}{0.48\linewidth}
\begin{center}
\includegraphics[trim=1mm 1mm 22mm 4mm, clip, width=65mm]{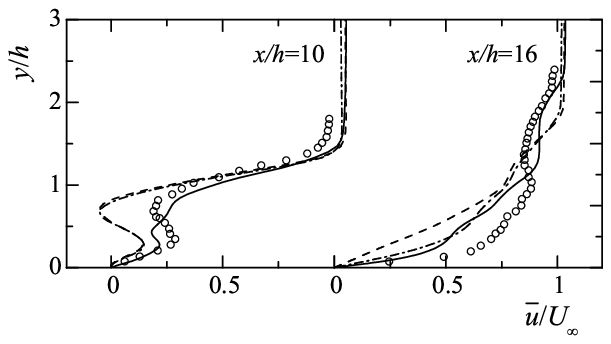} \\
(b) Second hill
\end{center}
\end{minipage}
\caption{Time-averaged distributions of streamwise velocity: 
---, $z/h=2.4 (L_z=4h)$; 
-\ -\ -, $z/h=3,9 (L_z=7h)$; \linebreak 
- $\cdot$ -, Yanaoka et al. ($\Rey=500$); $\circ$, Acarlar \& Smith ($\Rey=750$).}
\label{velocity}
\end{figure}
%------------------------------------------------------------------------------

%------------------------------------------------------------------------------
% Figure 8
%------------------------------------------------------------------------------
\begin{figure}
\begin{minipage}{0.48\linewidth}
\begin{center}
\includegraphics[trim=1mm 1mm 22mm 4mm, clip, width=65mm]{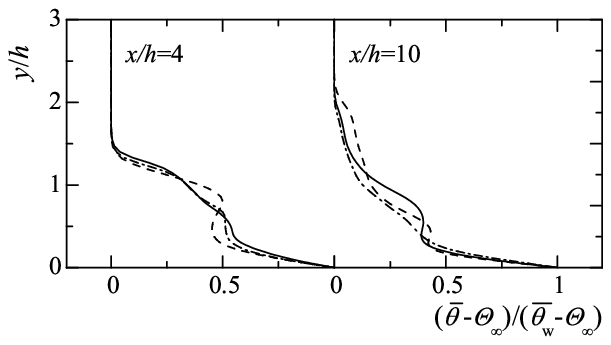} \\
(a) First hill
\end{center}
\end{minipage}
\hspace{0.02\linewidth}
\begin{minipage}{0.48\linewidth}
\begin{center}
\includegraphics[trim=1mm 1mm 22mm 4mm, clip, width=65mm]{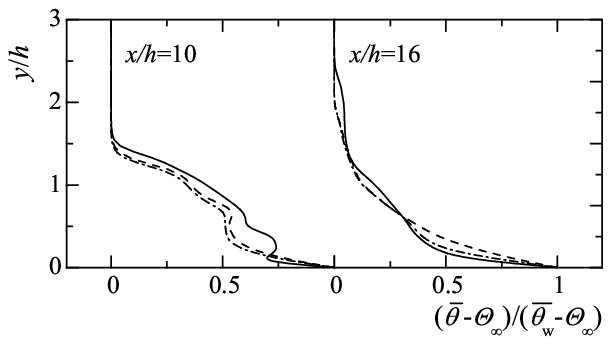} \\
(b) Second hill
\end{center}
\end{minipage}
\caption{Time-averaged distributions of temperature: 
---, $z/h=0.4 (L_z=4h)$; 
-\ -\ -, $z/h=0.4 (L_z=7h)$; 
- $\cdot$ -, Yanaoka et al. ($\Rey=500$).}
\label{temperature}
\end{figure}
%------------------------------------------------------------------------------

Figure \ref{vort_lz4} shows the streamwise vorticity ($\omega_x$) distribution 
in the $y$-$z$ plane for $L_z=4h$. 
Regions J to N are the distributions generated by the legs 
of hairpin vortices J to N, and region S is the distribution generated 
by the secondary vortex of hairpin vortex L. 
At $x/h=8$, high-vorticity region K appears below high-vorticity region J. 
At $x/h=10$ downstream, high-vorticity regions L and S
are generated by the legs of hairpin vortex L and the secondary vortex, 
respectively.

When $L_z=4h$, as can be seen from figures \ref{curv_lz4}(a) and \ref{vort_lz4}(a), 
both legs of hairpin vortex K are attracted to 
hairpin vortex J, which has high vorticity. 
As a result, the legs are spread. 
At this time, because the interference between the hairpin vortices 
becomes strong, the structure of hairpin vortex K 
changes significantly, and downstream, hairpin vortex K develops 
like hairpin vortex L with a small distance 
between the legs and high vorticity (figure \ref{vort_lz4}(b)). 
In hairpin vortex L, the distance between the legs narrows, 
and the legs significantly interfere with each other. 
Thus, the self-induction velocity between the legs increases, 
and Q2 ejection becomes strong. 
As a result, the head of hairpin vortex L rises sharply. 
Toward the downstream, because hairpin vortex M, 
whose vorticity weakens, is attracted to the developed secondary vortex S, 
the legs of hairpin vortex M spread. 
In hairpin vortex N, the shape of the head is close to that for $L_z=7h$. 
As the interference between adjacent hairpin vortices is weak for $L_z=7h$, 
the vortex structure and behaviour of the hairpin vortex in the wake 
do not significantly change like hairpin vortices K, L, and M for $L_z=4h$.

%++++++++++++++++++++++++++++++++++++++++++++++++++++++++++++++++++++++++++++++
\subsection{Mean properties}
%++++++++++++++++++++++++++++++++++++++++++++++++++++++++++++++++++++++++++++++

Figures \ref{velocity} and \ref{temperature} show time-averaged streamwise velocity 
and temperature distributions in the cross-sections of $z/h=3.9$ ($L_z=7h$), 
$z/h=2.4$ ($L_z=4h$), $z/h=0.4$ ($L_z=7h$), and $z/h=0.4$ ($L_z=4h$), 
where the hairpin vortices generated from the hills in the first 
and second rows respectively exist. 
The streamwise velocity is compared with the experimental values 
at $x/h=4$ and $x/h=10$ for $\Rey=750$ reported by \cite{Acarlar&Smith_1987a} 
and the calculation results at $x/h=4$ and $x/h=10$ for $\Rey=500$ 
previously reported by the present authors \citep{Yanaoka_et_al_2007b,Yanaoka_et_al_2008b}.
The temperature is compared with the results at $x/h=4$ and $x/h=10$ 
previously reported by the present authors \citep{Yanaoka_et_al_2007b,Yanaoka_et_al_2008b}. 
The positions of the calculation results shown 
in figures \ref{velocity}(b) and \ref{temperature}(b) 
coincide with the streamwise distance from the hill vertex 
in previously reported results 
\citep{Acarlar&Smith_1987a, Yanaoka_et_al_2007b, Yanaoka_et_al_2008b}. 
At $x/h=4$ in figure \ref{velocity}(a) and $x/h=10$ in figure \ref{velocity}(b), 
the maximum values generated by the legs of the hairpin vortex appear 
around $y/h=0.3$ in the distributions for $L_z=4h$ and $7h$, 
as is the case for the previously reported results \citep{Acarlar&Smith_1987a, Yanaoka_et_al_2007b, Yanaoka_et_al_2008b}. 
As both legs of the hairpin vortex develop, 
they approach the vicinity of the wall surface, 
and a secondary vortex is generated under the head. 
Therefore, the flows near the wall surface at $x/h=10$ and $x/h=16$ 
are faster than those at $x/h=4$ and $x/h=10$, respectively.

For $L_z=7h$ in figure \ref{velocity}(b), the distribution at $x/h=10$ 
well agrees with the previously reported result for a single hill 
\citep{Yanaoka_et_al_2007b, Yanaoka_et_al_2008b} 
but the distribution shape at $x/h=16$ is different from 
the previously reported shape. 
This result suggests that the hairpin vortices 
interfere with each other even for $L_z=7h$. 
In the distribution for $L_z=4h$ at $x/h=10$ in figure \ref{velocity}(b), 
the flow in the region from $y/h=0$ to 1.1 is faster than that in
the distribution for $L_z=7h$ and previously reported results 
\citep{Yanaoka_et_al_2007b, Yanaoka_et_al_2008b}. 
This is because the interference between the hairpin vortices narrows 
the legs of the hairpin vortex generated from the hills 
in the second row. 
In addition, the effect of contraction between the hills 
increases the flow velocity.

All temperature distributions in figure \ref{temperature} are high 
from the wall surface to $y/h=1$ 
because Q2 ejection and the rotation of the legs of the hairpin vortex 
lift the hot fluid near the wall. 
For $L_z=7h$ in figure \ref{temperature}(b), 
the distribution at $x/h=10$ well agrees with 
the previously reported result \citep{Yanaoka_et_al_2007b, Yanaoka_et_al_2008b}, 
similar to the streamwise velocity distribution, 
whereas the distribution at $x/h=16$ is slightly different from the previously reported result 
near the wall surface. 
The temperature distribution for $L_z=4h$ at $x/h=10$ 
in figure \ref{temperature}(b) shows higher values than 
that for $L_z=7h$ and the previously reported result 
\citep{Yanaoka_et_al_2007b, Yanaoka_et_al_2008b}. 
This is because the vorticity of the hairpin vortices generated 
from the hills in the second row increases 
due to the strong interference between the hairpin vortices, 
which increases the vertical self-induction velocity and 
transports the hot fluid to the mainstream side.

From these mean properties, 
it can be seen that hairpin vortices interfere with each other 
in a flow field where multiple hairpin vortices exist, 
and that the interference is weak for $L_z=7h$ but strong for $L_z=4h$.

%------------------------------------------------------------------------------
% Figure 9
%------------------------------------------------------------------------------
\begin{figure}
\begin{center}
\includegraphics[trim=3mm 4mm 5mm 5mm, clip, height=25mm]{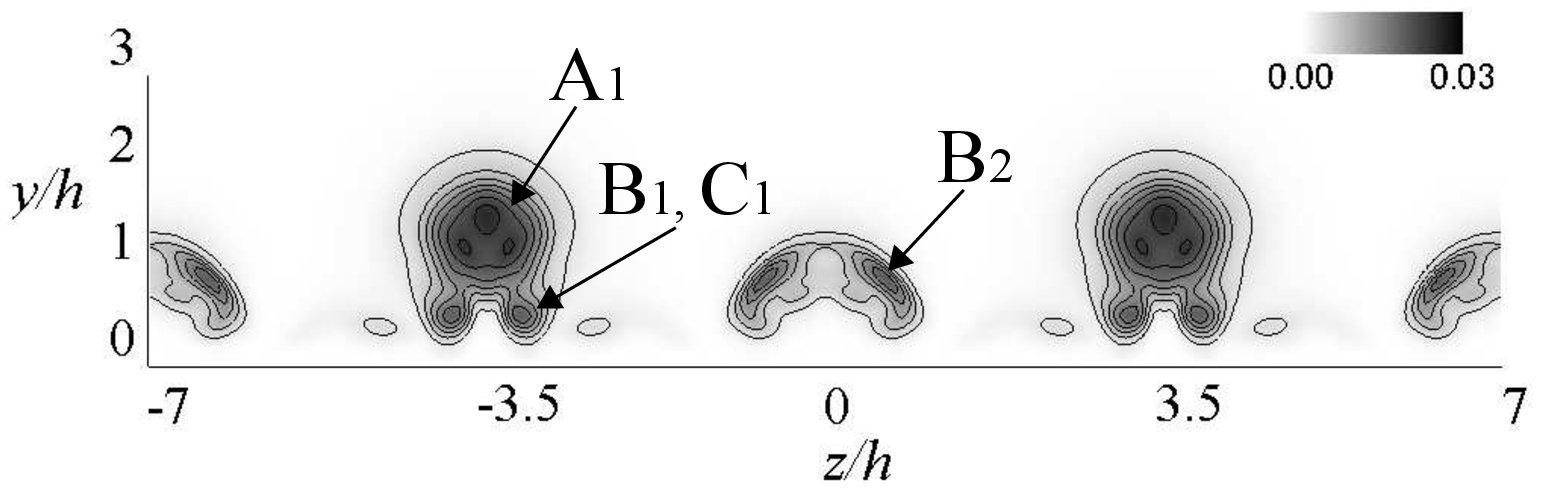} \\
(a) $x/h=10$ ($L_z=7h$) \\
\vspace*{1.0\baselineskip}
\includegraphics[trim=3mm 4mm 5mm 5mm, clip, height=25mm]{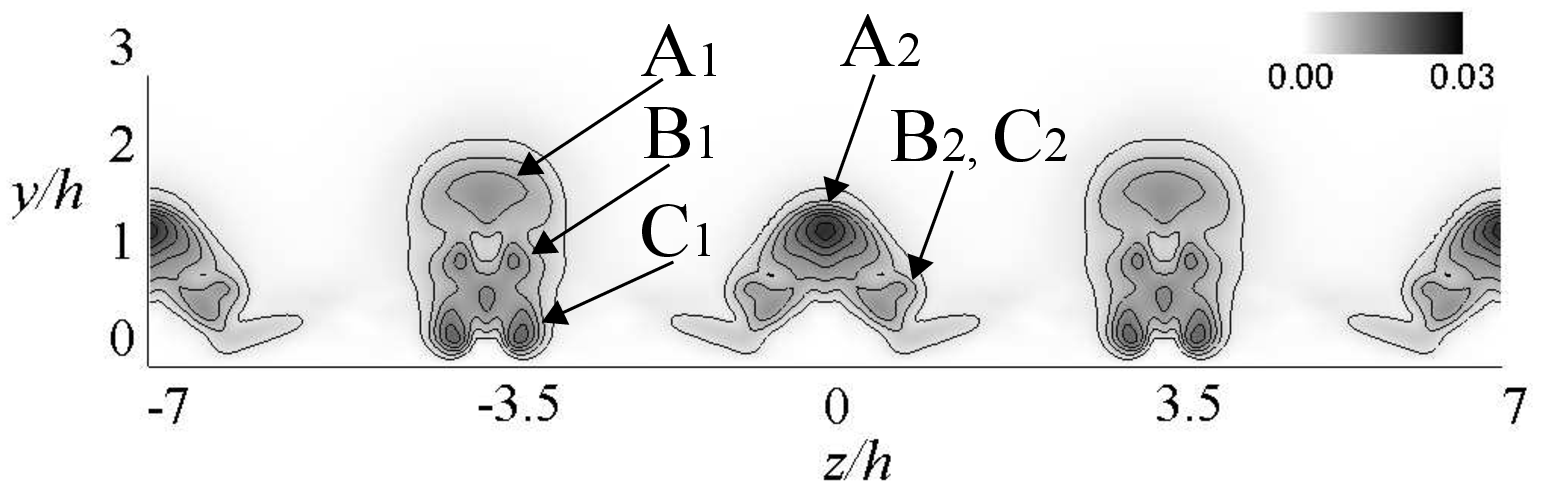} \\
(b) $x/h=15$ ($L_z=7h$) \\
\end{center}
\vspace*{1.0\baselineskip}
\begin{minipage}{0.48\linewidth}
\begin{center}
\includegraphics[trim=2mm 2mm 2mm 1mm, clip, height=25mm]{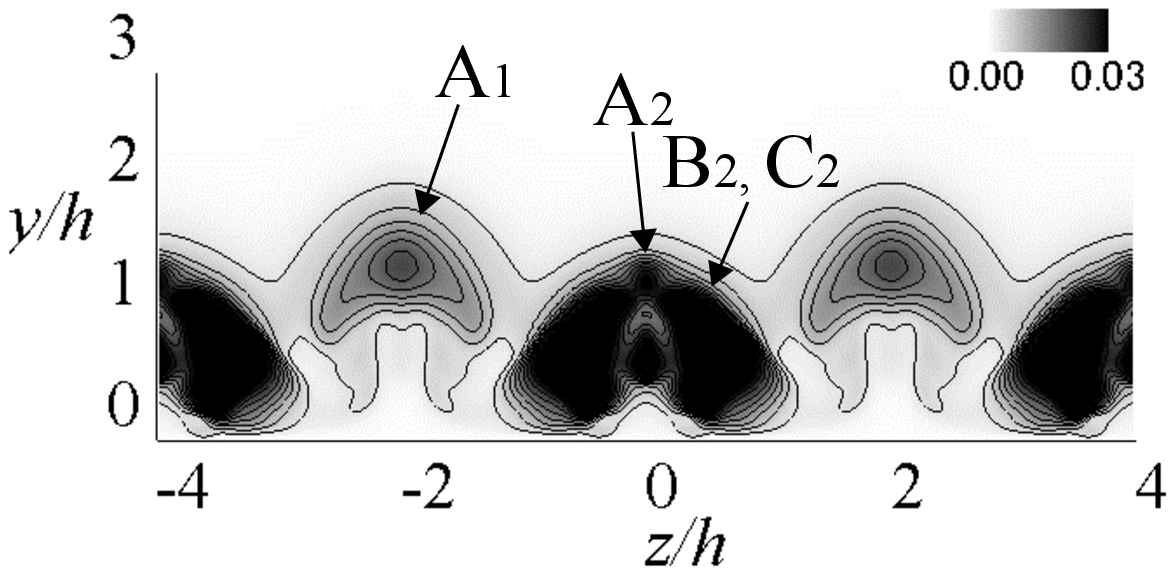} \\
(c) $x/h=10$ ($L_z=4h$)
\end{center}
\end{minipage}
\hspace{0.02\linewidth}
\begin{minipage}{0.48\linewidth}
\begin{center}
\includegraphics[trim=2mm 2mm 2mm 1mm, clip, height=25mm]{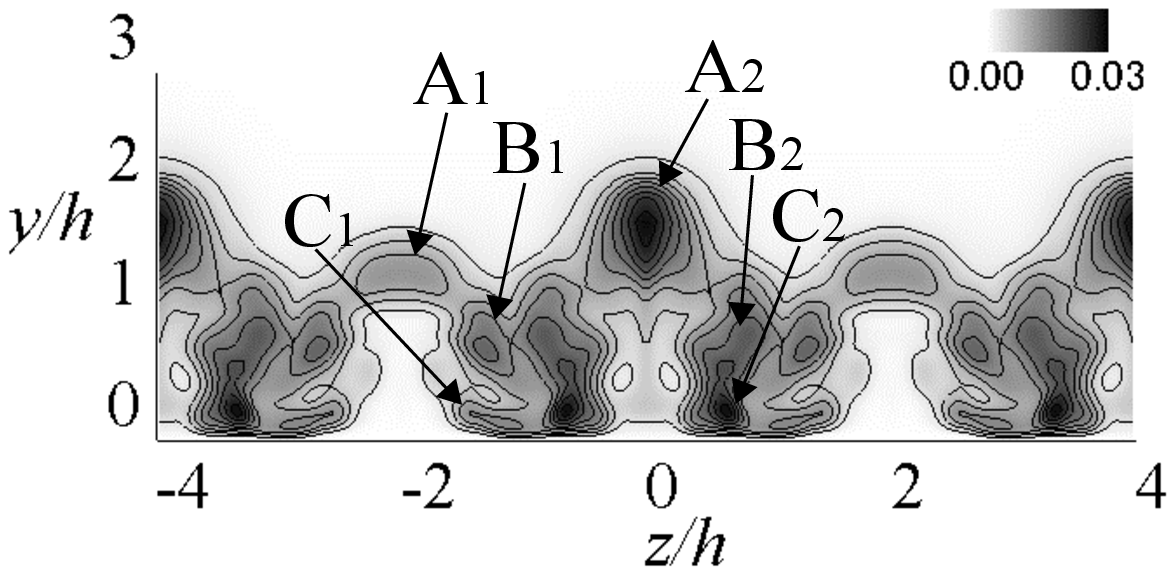} \\
(d) $x/h=15$ ($L_z=4h$)
\end{center}
\end{minipage}
\caption{Turbulence kinetic energy in $y$-$z$ plane at $x/h=10$ and $15$ 
(contour interval is 0.003 from 0 to 0.03).}
\label{k2}
\end{figure}
%------------------------------------------------------------------------------

%++++++++++++++++++++++++++++++++++++++++++++++++++++++++++++++++++++++++++++++
\subsection{Turbulence characteristics}
%++++++++++++++++++++++++++++++++++++++++++++++++++++++++++++++++++++++++++++++

Figure \ref{k2} shows the contours of the time-averaged turbulence kinetic energy 
in the $y$-$z$ plane at $x/h=10$ and 15 for $L_z=7h$ and $4h$. 
Here, high-turbulence regions caused by the head and legs of the hairpin vortex 
and the secondary vortex are labelled A, B, and C, respectively. 
In addition, the turbulences caused by the hairpin vortices 
generated from the hills in the first and second rows are distinguished by 
subscripts 1 and 2, respectively. 
At $x/h=10$ for $L_z=7h$, the head of the hairpin vortex 
generated from the hills in the first row forms high-turbulence region A$_1$ 
around $z/h=\pm 3.5$. 
Turbulence regions B$_1$ and C$_1$ near the wall surface are generated by 
the two legs of the hairpin vortex and the secondary vortex, respectively. 
Around $z/h=0$, turbulence region B$_2$ is generated by the legs of the hairpin vortex. 
The above trend is the same at $x/h=15$ downstream. 
Because interference between hairpin vortices is weak for $L_z=7h$, 
such turbulence distributions generated by hairpin vortices are similar to 
previously reported results 
\citep{Gretta&Smith_1993, Dong&Meng_2004, Yanaoka_et_al_2007b, Yanaoka_et_al_2008b}.

At $x/h=10$ for $L_z=4h$, the head of the hairpin vortex 
generated from the hill in the first row forms turbulence region A1 
around $z/h=\pm2$. 
Near $z/h=0$, regions A2, B2, and C2 have higher turbulence than those for $L_z=7h$. 
This result indicates that the interference between the hairpin vortices remarkably 
strengthens the hairpin vortex generated from the hill in the second row 
and that the secondary vortex develops into a strong vortex. 
Furthermore, because the distance between the legs of the hairpin vortex 
is narrowed in this region, 
the strong interference between the legs 
is considered to be a factor in such high turbulence. 
At $x/h=15$ downstream, because the legs of the hairpin vortex 
at $z/h=\pm2$ are widened by the neighbour hairpin vortex, 
high-turbulence regions B1 and C1 respectively 
generated by the legs of the hairpin vortex and the secondary vortex, 
approach regions B2 and C2. 
Therefore, the high-turbulence region spreads 
in the spanwise direction near the wall surface.

%--------------------------------------------------------------------------
% Figure 10
%--------------------------------------------------------------------------
\begin{figure}
\begin{minipage}{0.48\linewidth}
\begin{center}
\includegraphics[trim=1mm 1mm 0mm 1mm, clip, width=65mm]{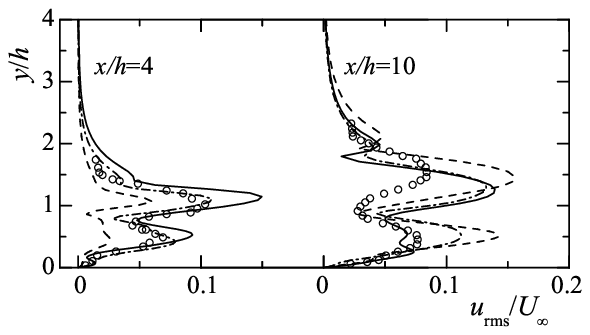} \\
(a) First hill \\
\end{center}
\end{minipage}
\hspace{0.02\linewidth}
\begin{minipage}{0.48\linewidth}
\begin{center}
\includegraphics[trim=1mm 1mm 0mm 1mm, clip, width=65mm]{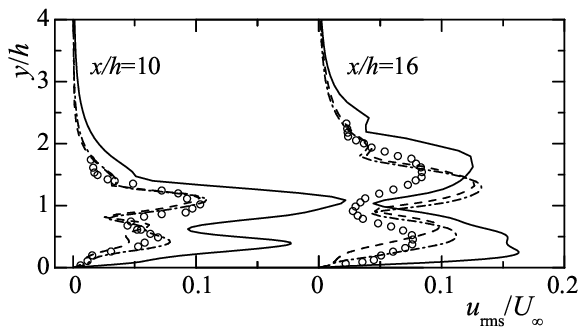} \\
(b) Second hill \\
\end{center}
\end{minipage}
\caption{Turbulence intensity distributions of streamwise velocity fluctuation: 
---, $L_{z}=4h$; - - -, $L_{z}=7h$; 
- $\cdot$ -, Yanaoka et al. ($\Rey=500$); $\circ$, Acarlar \& Smith ($\Rey=750$).}
\label{urms}
\end{figure}
%------------------------------------------------------------------------------

Figure \ref{urms} compares the turbulence intensity distributions of 
the streamwise velocity fluctuations in the cross-sections 
of $z/h=3.9$ ($L_z=7h$), $z/h=2.4$ ($L_z=4h$), 
$z/h=0.4$ ($L_z=7h$), and $z/h=0.4$ ($L_z=4h$) 
with the previously reported results 
\citep{Acarlar&Smith_1987a, Yanaoka_et_al_2007b, Yanaoka_et_al_2008b}. 
The plane position is the same as that in figures \ref{velocity} and \ref{temperature}. 
The position of the calculation result shown in figure \ref{urms}(b) 
coincides with the streamwise direction distance from the hill vertex 
in the previously reported results 
\citep{Acarlar&Smith_1987a, Yanaoka_et_al_2007b, Yanaoka_et_al_2008b}. 
In the distributions of $L_z=7h$ and $4h$ at $x/h=4$ 
shown in figure \ref{urms}(a), two maxima are generated around $y/h=1.1$ and 0.5 
by the head and the two legs of the hairpin vortex, respectively. 
At $x/h=10$ downstream, the maximum value on the mainstream side 
occurs at a high position as the head rises 
due to the development of the hairpin vortex. 
Near the wall surface, the legs of the hairpin vortex approach 
the wall surface and the secondary vortex develops, 
so high turbulence is maintained. 
In the distributions of $L_z=7h$ and $4h$ in figure \ref{urms}(b), 
two maxima are also generated by the head and the two legs of the hairpin vortex 
as in figure \ref{urms}(a).

In the distribution for $L_z=4h$ at $x/h=4$ in figure \ref{urms}(a), 
higher turbulence occurs near $y/h=1.1$ than that for $L_z=7h$ 
because the hairpin vortex generated from the hill in the first row interacts with 
the flow entering between the hills in the first row. 
In the distribution for $L_z=4h$ at $x/h=10$ in figure \ref{urms}(b), 
the turbulence is significantly higher than that for $L_z=7h$, 
previously reported results 
\citep{Acarlar&Smith_1987a, Yanaoka_et_al_2007b, Yanaoka_et_al_2008b}, 
and the result for $L_z=4h$ at $x/h=10$ in figure \ref{urms}(a). 
This is because the strong interference between the hairpin vortices 
strengthens the interference between the legs of the hairpin vortices 
generated from the hills in the second row, 
and Q2 ejection and the secondary vortex also strengthen. 
In the distribution for $L_z=4h$ at $x/h=16$ downstream, 
the turbulence due to the head occurs at a higher position than 
that for $L_z=7h$ and the previously reported results 
\citep{Acarlar&Smith_1987a, Yanaoka_et_al_2007b, Yanaoka_et_al_2008b}. 
The turbulence is high near the wall surface 
due to the two legs of the strengthened hairpin vortex 
and the secondary vortex.

Next, figure \ref{k} shows the streamwise variation of 
the area- and time-averaged turbulence kinetic energy $k$ 
in the $y$-$z$ plane. 
Because the hairpin vortex generated from the hill in the first row develops, 
the value of $k$ begins to increase from $x/h=2$ for $L_z=4h$ 
and $x/h=3$ for $L_z=7h$. 
Downstream from the vicinity of $x/h=8$ for $L_z=4h$ and $x/h=9$ for $L_z=7h$, 
the hairpin vortices generated from the hills in the first and second rows 
begin to interfere with each other, 
so the value of $k$ further increases. 
Near $x/h=17$ for $L_z=4h$ and $x/h=14$ for $L_z=7h$, 
the value of $k$ decreases because the hairpin vortex decays 
and the interference between the hairpin vortices begins to weaken. 
The maximum value of $k$ around $x/h=10$ for $L_z=4h$ reaches 4.2 times 
the maximum value around $x/h=12$ for $L_z=7h$, 
and the interference between hairpin vortices is remarkably strong for $L_z=4h$.

%++++++++++++++++++++++++++++++++++++++++++++++++++++++++++++++++++++++++++++++
\subsection{Heat transport property}
%++++++++++++++++++++++++++++++++++++++++++++++++++++++++++++++++++++++++++++++

%------------------------------------------------------------------------------
% Figure 11
%------------------------------------------------------------------------------
\begin{figure}
\begin{center}
\includegraphics[trim=1mm 0mm 3mm 3mm, clip, width=65mm]{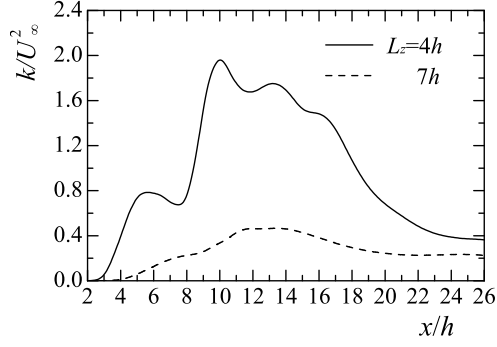} \\
\end{center}
\caption{Area- and time-averaged distributions of turbulence kinetic energy 
in $y$-$z$ plane.} 
\label{k}
\end{figure}
%------------------------------------------------------------------------------

%--------------------------------------------------------------------------
% Figure 12
%--------------------------------------------------------------------------
\begin{figure}
\begin{center}
\includegraphics[trim=21mm 5mm 19mm 12mm, clip, height=25mm]{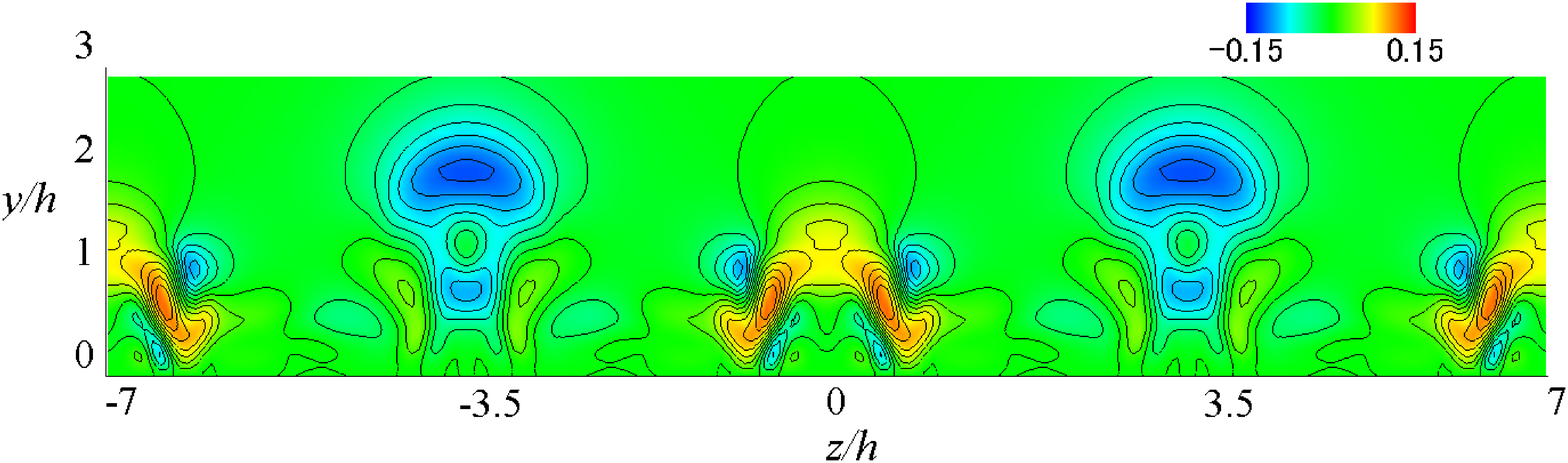} \\
\includegraphics[trim=1mm 3mm 6mm 6mm, clip, height=25mm]{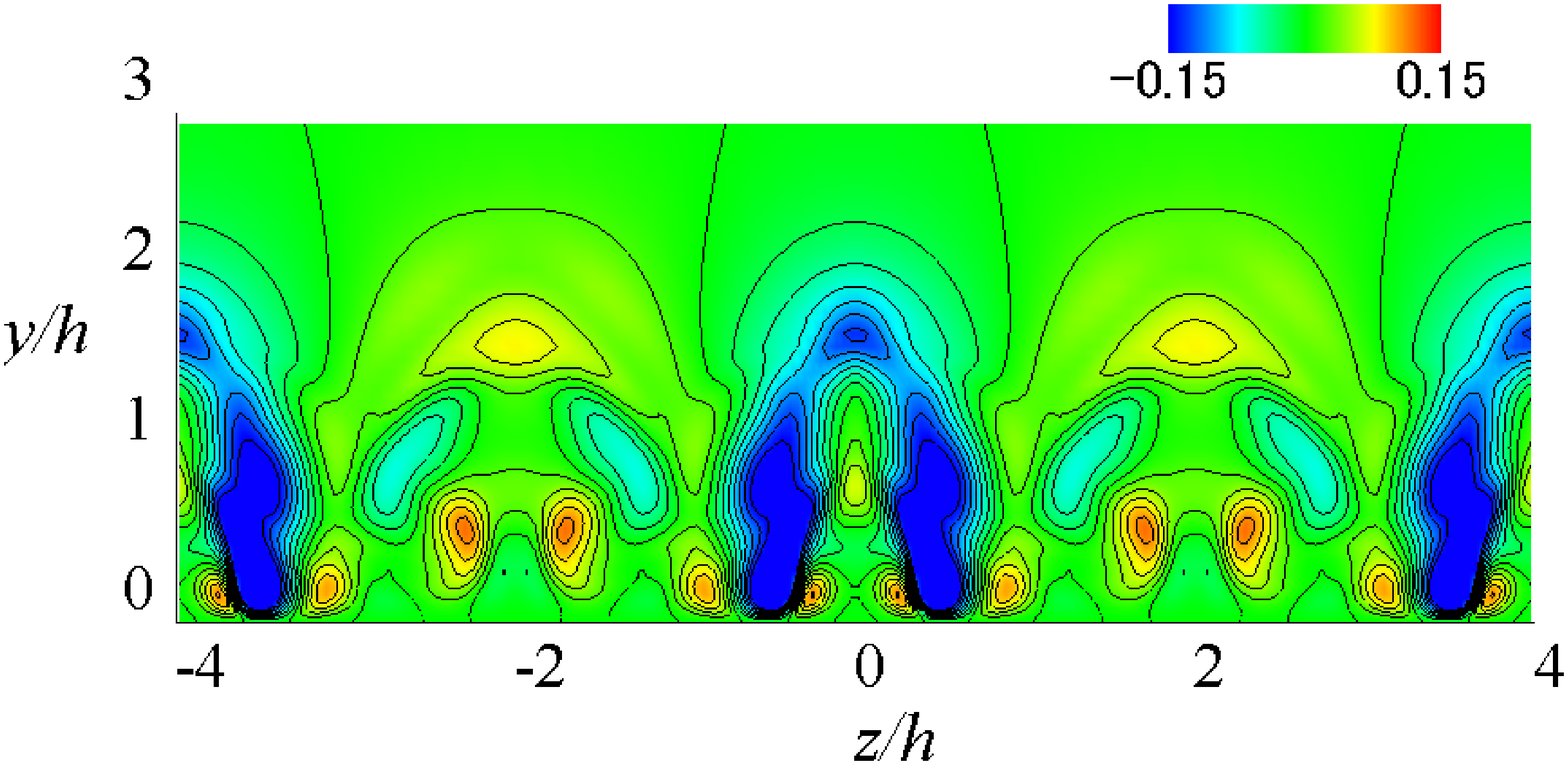} \\
(a) Vertical velocity fluctuation (contour interval is $0.02$ from $-0.15$ to $0.15$)
\end{center}
\vspace*{1.0\baselineskip}
\begin{center}
\includegraphics[trim=21mm 5mm 16mm 12mm, clip, height=25mm]{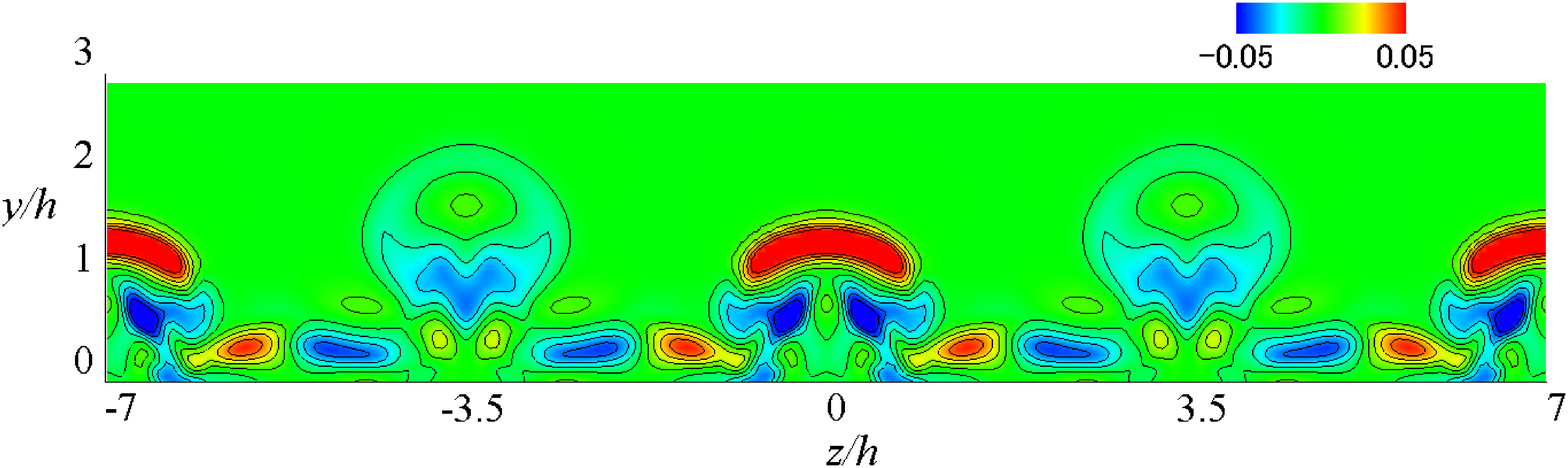} \\
\includegraphics[trim=1mm 3mm 3mm 6mm, clip, height=25mm]{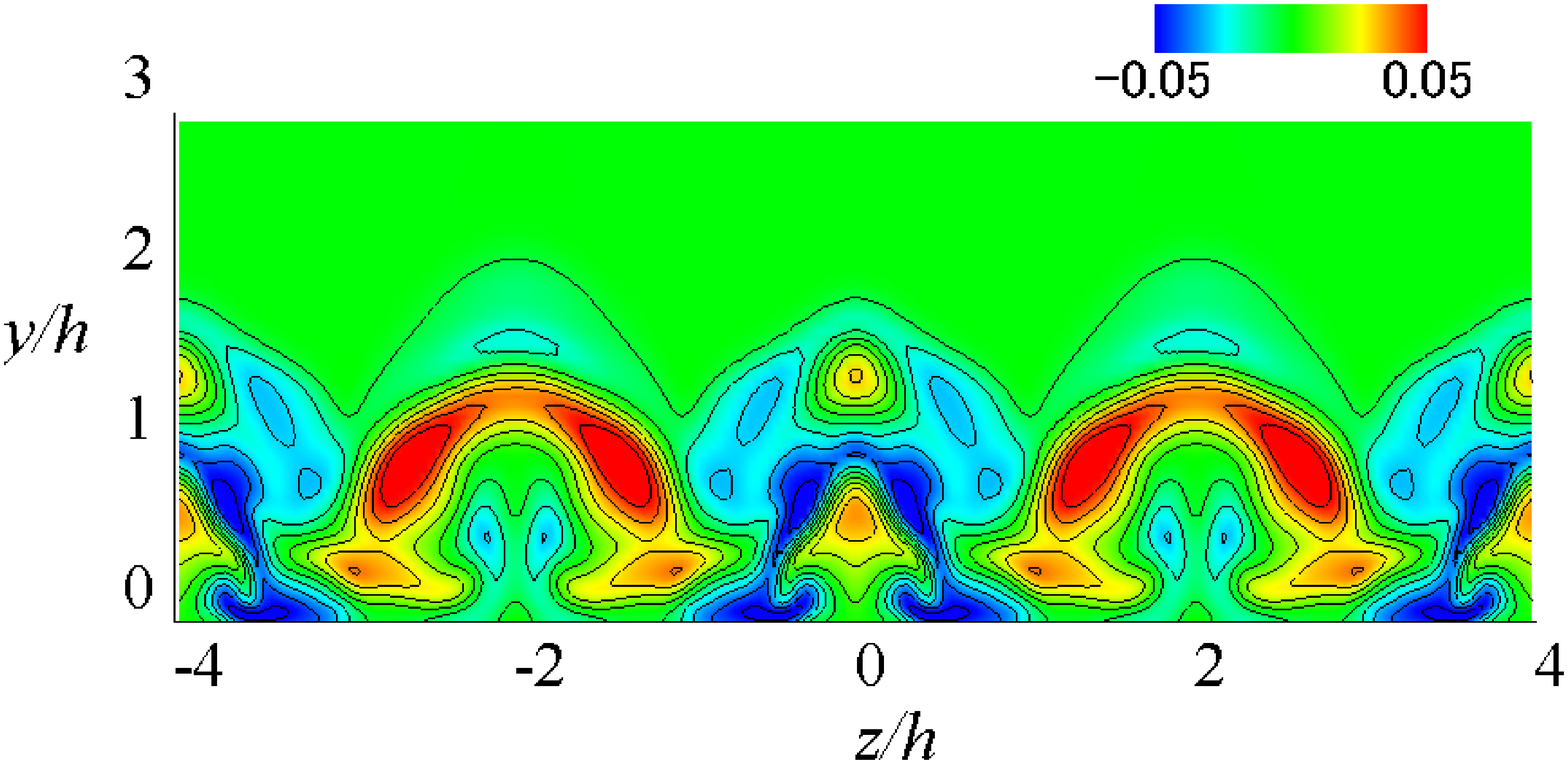} \\
(b) Temperature fluctuation (contour interval is $0.01$ from $-0.05$ to $0.05$)
\end{center}
\caption{Vertical velocity and temperature fluctuations of $L_z=7h$ (top) 
and $L_z=4h$ (bottom) at $x/h=12$ in $y$-$z$ plane.}
\label{fluc}
\end{figure}
%--------------------------------------------------------------------------

To investigate the heat transport in the flow field, figure \ref{fluc} 
shows the contours of the instantaneous vertical velocity 
and temperature fluctuations in the $y$-$z$ plane at $x/h=12$, 
where high-turbulence kinetic energy was found for $L_z=4h$ and $7h$ 
in figure \ref{k}. 
For $L_z=7h$, the vertical velocity fluctuation 
around $y/h=$2 and $z/h=\pm 3.5$ is negative. 
This fluctuation is mainly due to the Q4 sweep induced 
around the head of the hairpin vortex. 
Because the low-temperature fluid of the mainstream is transported 
toward the wall surface by the Q4 sweep, the temperature fluctuation 
becomes negative around $y/h=1$ in this region. 
The Q2 ejection by the head of the hairpin vortex 
and the rotating flow of the legs occur. 
Therefore, from $y/h=0$ to $y/h=1$ around $z/h=\pm0.4$, 
the vertical velocity fluctuation becomes positive. 
At that time, because the high-temperature fluid is lifted by upward flow, 
the temperature fluctuation around $y/h=1.3$ becomes 
an arch-shaped positive distribution centred on $z/h=0$. 
Near the wall surface around $z/h=\pm0.3$, 
the vertical velocity and temperature fluctuations become negative 
due to the rotation of the secondary vortex.

%--------------------------------------------------------------------------
% Figure 13
%--------------------------------------------------------------------------
\begin{figure}
\begin{minipage}{0.48\linewidth}
\begin{center}
\includegraphics[trim=0mm 0mm 10mm 0mm, clip, width=65mm]{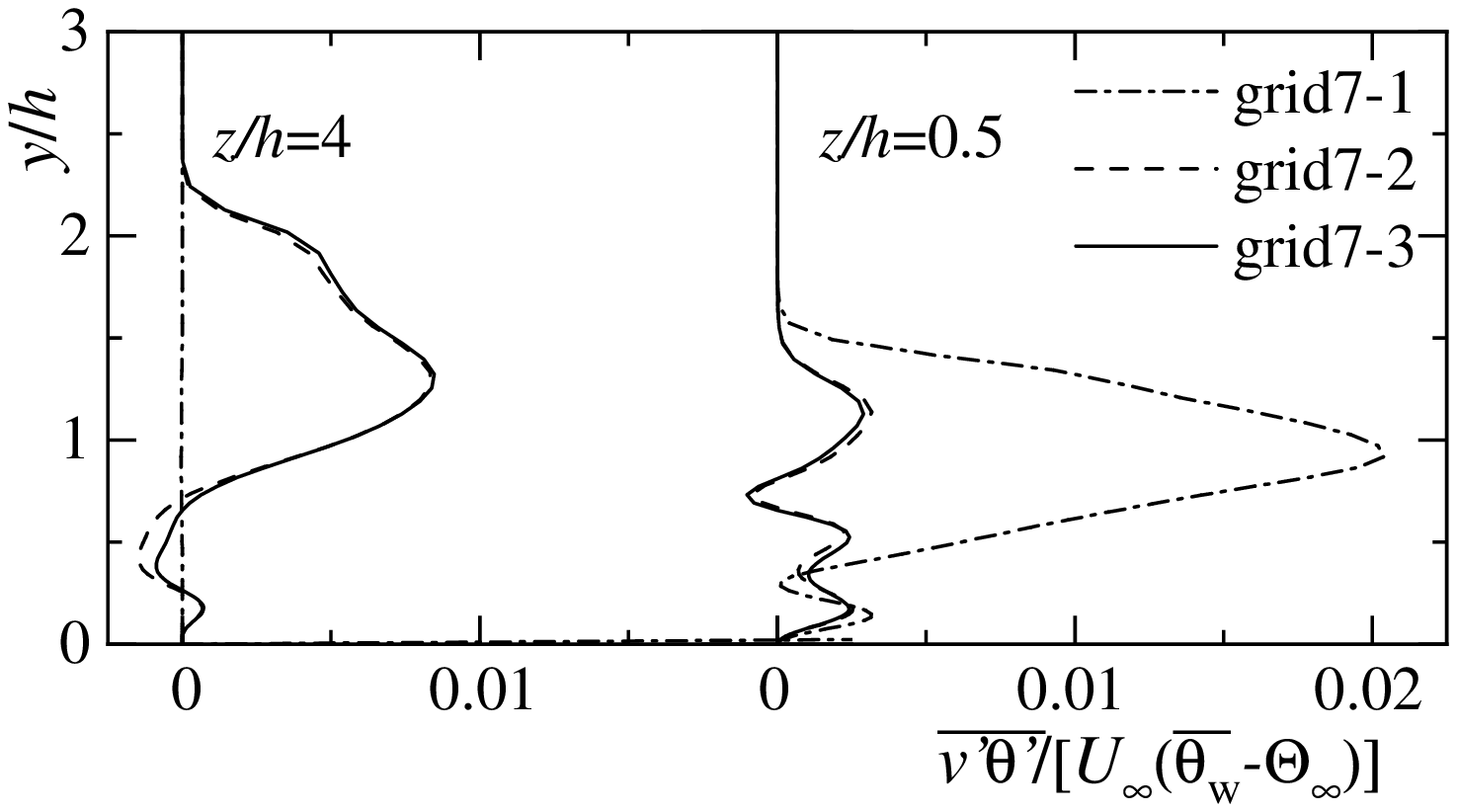} \\
(a) $L_z=7h$ \\
\end{center}
\end{minipage}
\hspace{0.02\linewidth}
\begin{minipage}{0.48\linewidth}
\begin{center}
\includegraphics[trim=0mm 0mm 10mm 0mm, clip, width=65mm]{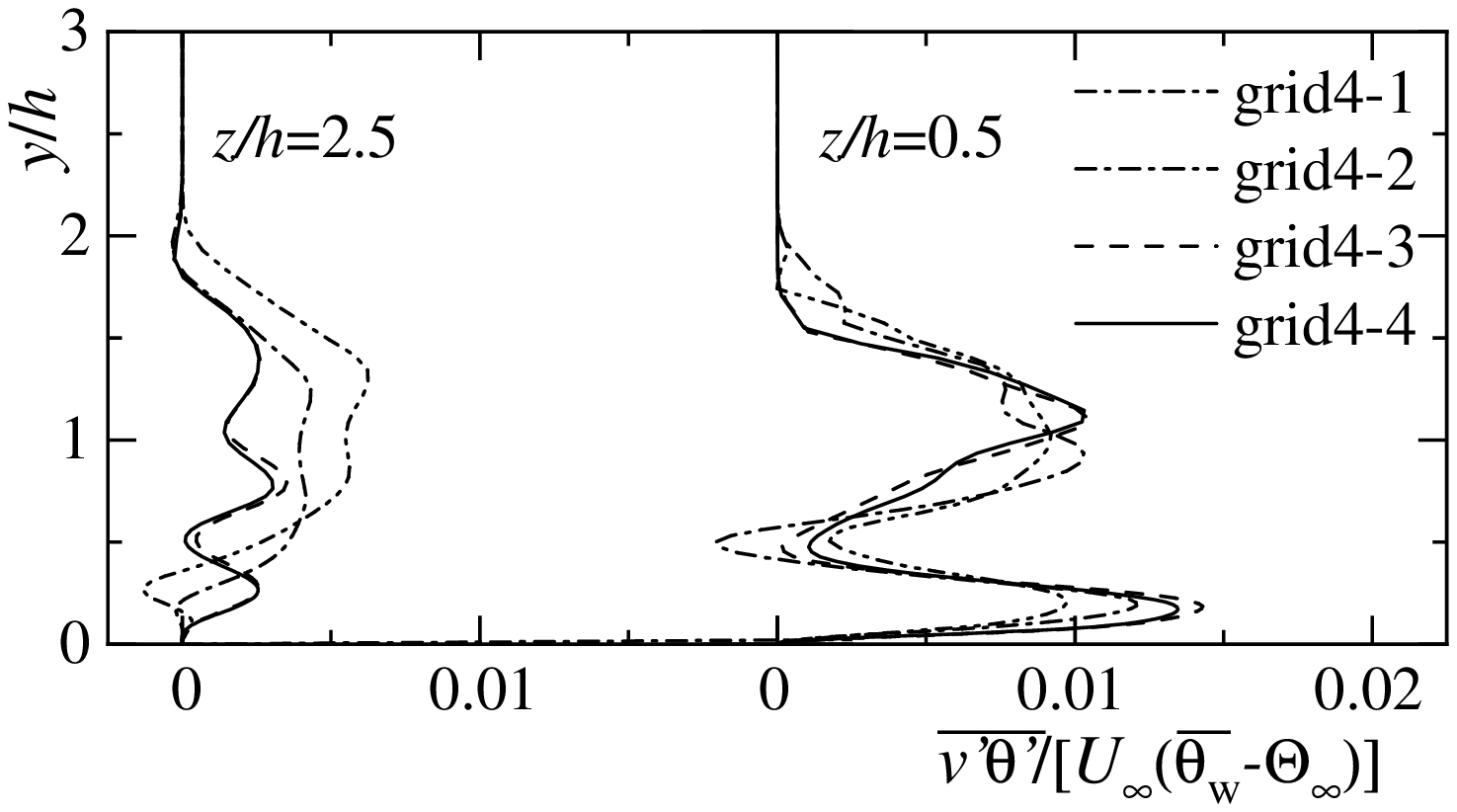} \\
(b) $L_z=4h$ \\
\end{center}
\end{minipage}
\caption{Distributions of turbulent heat flux at $x/h=12$.}
\label{vt05}
\end{figure}
%--------------------------------------------------------------------------

The trend of $L_z=4h$ is similar to that of $L_z=7h$. 
For $L_z=4h$, the hairpin vortex generated from the hill 
in the second row is strengthened by the interference with 
the adjacent hairpin vortex, 
and the secondary vortex strengthens accordingly. 
Therefore, strong fluctuations occur in the entire flow field, 
compared with the result for $L_z=7h$.

Next, figure \ref{vt05} shows the distributions of the turbulent heat flux 
$\overline{v'\theta'}$ at the planes of $z/h=4$ ($L_z=7h$), 
$z/h=2.5$ ($L_z=4h$), $z/h=0.5$ ($L_z=7h$), and $z/h=0.5$ ($L_z=4h$), 
where the legs of the hairpin vortex generated from the hill 
in the first and second rows exist. 
In this figure, the results obtained using each grid point are compared. 
The maximum value of $\overline{v'\theta'}$ on the mainstream side 
of each distribution is generated by the head of the hairpin vortex. 
In addition, the maximum value of $\overline{v'\theta'}$ appears 
near the wall surface due to the legs of the hairpin vortex 
and the secondary vortex. 
For $L_z=7h$ (grid7-3), the maximum regions are around $z/h=4$, $y/h=0.2$ 
and $z/h=0.5$, $y/h=0.2$, 0.5. 
For $L_z=4h$ (grid4-4), the regions are around $z/h=2.5$, $y/h=0.3$ 
and $z/h=0.5$, $y/h=0.2$.
The hairpin vortex generated from the hill in the second row 
and the secondary vortex are strengthened by the strong interference 
between the hairpin vortices. 
Therefore, in the distribution for $L_z=4h$ at $z/h=0.5$, 
$\overline{v'\theta'}$ is higher than that for $L_z=7h$, 
and heat transport is active.

There are differences in the results obtained with grid7-1 and grid7-2 for $L_z=7h$ 
and those obtained with grid4-1, grid4-2, and grid4-3 for $L_z=4h$. 
In contrast, 
the distributions obtained with grid7-2 and grid7-3 for $L_z=7h$ 
and those obtained with grid4-3 and grid4-4 for $L_z=4h$ well agree. 
Therefore, grid7-1 for $L_z=7h$ 
and grid4-1 and grid4-2 for $L_z=4h$ are coarse grids, 
and it can be said that the grid resolution is low.

\begin{figure}
\begin{minipage}{0.48\linewidth}
\begin{center}
\includegraphics[trim=0mm 0mm 3mm 3mm, clip, width=65mm]{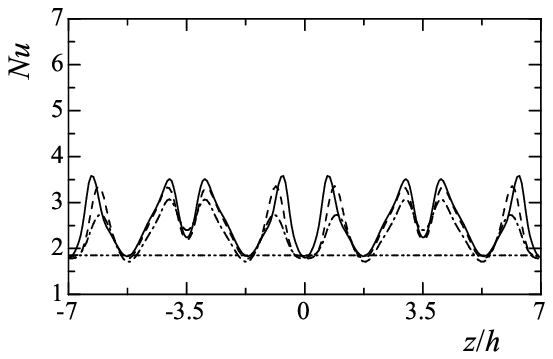} \\
(a) $L_z=7h$
\end{center}
\end{minipage}
\hspace{0.02\linewidth}
\begin{minipage}{0.48\linewidth}
\begin{center}
\includegraphics[trim=0mm 0mm 3mm 3mm, clip, width=65mm]{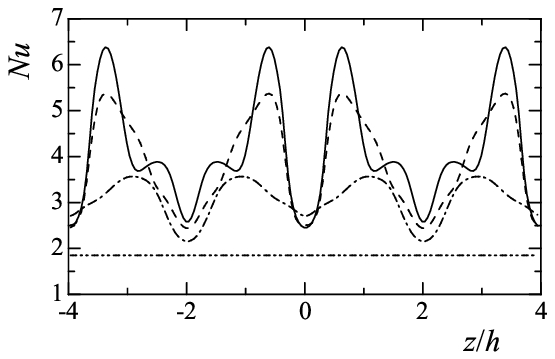} \\
(b) $L_z=4h$
\end{center}
\end{minipage}
\caption{Instantaneous Nusselt number distributions: 
---, $x/h=12$; - - -, $x/h=15$; 
- $\cdot$ -, $x/h=18$; 
- $\cdot$ $\cdot$ -, similarity solution at $x/h=12$.}
\label{nu_z}
\end{figure}

\begin{figure}
\begin{center}
\includegraphics[trim=0mm 0mm 0mm 4mm, clip, width=65mm]{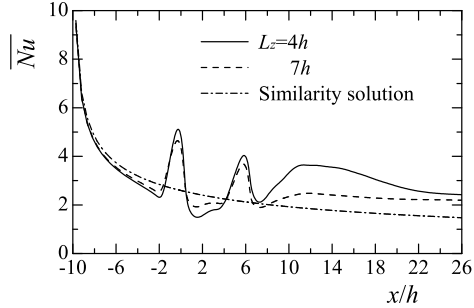} \\
\end{center}
\vspace*{-0.5\baselineskip}
\caption{Area- and time-averaged Nusselt number distributions.}
\label{nu}
\end{figure}

Next, figure \ref{nu_z} shows the variation 
of the instantaneous Nusselt number $Nu$ in the spanwise direction. 
Figure \ref{nu} shows the time- and spanwise-averaged Nusselt number $\overline{Nu}$. 
The Nusselt number is defined as $Nu=\alpha h/\lambda$, 
where $\alpha$ is the local heat transfer coefficient, 
$\lambda$ is the thermal conductivity, and $q_w$ is the heat flux at the wall. 
In figures \ref{nu_z} and \ref{nu}, the calculation results are compared 
with a similarity solution for the laminar boundary layer 
\citep{Lighthill_1950}, as in the previous studies \citep{Yanaoka_et_al_2007b,Yanaoka_et_al_2008b}. 
Around $z/h=\pm 4$, $\pm 3$, and $\pm 0.7$ for $L_z=7h$, $Nu$ is high 
because heat transport becomes active due to the flow of 
the legs of the hairpin vortex and the secondary vortex. 
In the distribution for $L_z=4h$, $Nu$ becomes maximum 
around $z/h=\pm 1.5$, $\pm 2.5$, and $\pm 0.6$ 
due to the same factors as those for $L_z=7h$. 
As can be seen from figure \ref{curv_lz4}(a), 
the maximum values around $z/h=\pm 0.6$ for $L_z=4h$ are strongly 
influenced by the secondary vortex for the distributions of $x/h=12$ 
and by the legs of the hairpin vortex for $x/h=15$.

For $L_z=4h$, $Nu$ is very high around $z/h=\pm 0.6$ 
compared to $Nu$ for $L_z=7h$. 
This is because the hairpin vortices interfere with each other. 
As the turbulence is increased by the legs of the strongly 
developed hairpin vortex and the secondary vortex, 
heat transport near the wall surface becomes active. 
In addition, $Nu$ is high at the plane of $x/h=12$, 
where the interference between hairpin vortices is strong. 
However, because the interference between hairpin vortices weakens 
downstream and heat transport declines, $Nu$ decreases. 
$Nu$ for $L_z=4h$ is higher over the entire region in the spanwise direction 
than $Nu$ for $L_z=7h$. 
As a result, $\overline{Nu}$ for $L_z=4h$ in figure \ref{nu} shows 
a higher value from $x/h=8$ to 20 than the distribution for $L_z=7h$ 
and the similarity solution. 
Therefore, it is clear that the heat transfer coefficient for $L_z=4h$ 
is high over a wide range of the wake.

%--------------------------------------------------------------------------
% Figure 14
%--------------------------------------------------------------------------
%--------------------------------------------------------------------------

%--------------------------------------------------------------------------
% Figure 15
%--------------------------------------------------------------------------
%--------------------------------------------------------------------------

%##############################################################################
\section{Conclusions}
%##############################################################################

We performed a numerical simulation of the interference 
and heat transfer between hairpin vortices formed in the wake 
behind staggered hills in a laminar boundary layer. 
The obtained findings are summarized as follows.

Hairpin vortices are periodically shed in the wake of a row of hills, 
causing interference between the hairpin vortices. 
As the spanwise distance between the hills decreases, 
the interference increases and the hairpin vortex becomes strong. 
Then, because the distance between the legs of the hairpin vortex becomes narrow, 
the interference between the legs and Q2 ejection becomes strong. 
As a result, the head of the hairpin vortex rises sharply.

When the hill spacing decreases, 
the turbulence caused by the head and legs of the hairpin vortex 
generated from the hill in the second row increases remarkably. 
In addition, the secondary vortex also generates turbulence. 
Downstream, the hairpin vortex and secondary vortex are attracted 
to adjacent hairpin vortices, causing widespread high turbulence 
in the spanwise direction near the wall surface.

Regardless of hill spacing, the Q2 ejection and Q4 sweep 
due to the hairpin vortex occur, 
and the secondary vortex forms around the hairpin vortex, 
activating heat transport and increasing the heat transfer coefficient 
in the wake. 
When the hill spacing becomes narrower, 
the interference between the hairpin vortices strengthens 
the legs and the secondary vortex, and heat transport 
near the wall surface becomes very active. 
In addition, the heat transfer increases over a wide range of the wake 
because the legs of hairpin vortices flowing downstream are spread 
in the spanwise direction.

%##############################################################################
%\section*{Acknowledgment}
%##############################################################################

\vspace*{1.0\baselineskip}
\noindent
{\bf Acknowledgements.}
The numerical results in this research were obtained 
using supercomputing resources at the Cyberscience Center, Tohoku University. 
This research did not receive any specific grant from funding agencies 
in the public, commercial, or not-for-profit sectors. 
We would like to express our gratitude to Associate Professor Yosuke Suenaga 
of Iwate University for his support of our laboratory. 
The authors wish to acknowledge the time and effort of everyone involved in this study.

\vspace*{1.0\baselineskip}
\noindent
{\bf Declaration of interests.}
The authors report no conflicts of interest.

\vspace*{1.0\baselineskip}
\noindent
{\bf Author ORCID.} \\
H. Yanaoka \url{https://orcid.org/0000-0002-4875-8174}.

\vspace*{1.0\baselineskip}
\noindent
{\bf Author contributions.}
H. Y. conceived and planned the research 
and developed the calculation method and numerical codes. 
T. Y. performed the simulations. 
All authors contributed equally to analysing data, reaching conclusions, 
and writing the paper.

%##############################################################################
%\section*{References}
%##############################################################################

\bibliographystyle{arXiv_elsarticle-harv}
\bibliography{reference_turbulence_bibfile}

\end{document}